\documentclass[10pt,twocolumn,twoside]{IEEEtran}
\usepackage{graphicx}
\usepackage{amsmath}
\usepackage{array}
\newcommand{\PreserveBackslash}[1]{\let\temp=\\#1\let\\=\temp}
\newcolumntype{C}[1]{>{\PreserveBackslash\centering}p{#1}}
\newcolumntype{R}[1]{>{\PreserveBackslash\raggedleft}p{#1}}
\newcolumntype{L}[1]{>{\PreserveBackslash\raggedright}p{#1}}
\usepackage[usenames]{color}
\usepackage{colortbl,booktabs}
\usepackage{tabularx}
\usepackage{multicol}
\usepackage{booktabs}
\usepackage[Symbol]{upgreek}
\usepackage{subfigure}
\usepackage{bm}
\usepackage{cite}
\usepackage{balance}
\usepackage[boxed]{algorithm2e}

\usepackage{threeparttable}
\usepackage{amsmath}
\usepackage{dcolumn}
\usepackage{multirow}
\usepackage{booktabs}
\usepackage{amsfonts}
\newcolumntype{d}[1]{D{.}{.}{#1}}

\def \qed {\hfill \vrule height6pt width 6pt depth 0pt}

\begin{document}

\bibliographystyle{IEEEtran} 
\title{Near-Optimal Linear Precoding with Low Complexity for Massive MIMO}

\author{Linglong Dai,~\IEEEmembership{Senior Member,~IEEE}, Xinyu Gao,~\IEEEmembership{Student Member,~IEEE}, Shuangfeng Han,~\IEEEmembership{Member,~IEEE},  \\ Chih-Lin I,~\IEEEmembership{Senior Member,~IEEE}, and Zhaocheng Wang,~\IEEEmembership{Senior Member,~IEEE}

\thanks{L. Dai, X. Gao, and Z. Wang are with the Tsinghua National Laboratory
for Information Science and Technology (TNList), Department of Electronic Engineering, Beijing 100084, China (e-mail: daill@tsinghua.edu.cn).}
\thanks{S. Han and C. I are with the Green Communication Research Center, China Mobile Research
Institute, Beijing 100053, China (e-mail: hanshuangfeng@chinamobile.com).}
\thanks{This work was supported by National Key Basic Research Program of
China (Grant No. 2013CB329203), National Natural Science Foundation
of China (Grant Nos. 61271266 and 61201185), and Science and Technology Foundation for Beijing Outstanding Doctoral Dissertation (Grant No. 2012T50093).}}


\maketitle
\begin{abstract}
Linear precoding techniques can achieve near-optimal capacity due to the special channel property in downlink massive MIMO systems, but involve high complexity since complicated matrix inversion of large size is required. In this paper, we propose a low-complexity linear precoding scheme based on the Gauss-Seidel (GS) method. The proposed scheme can achieve the capacity-approaching performance of the classical linear precoding schemes in an iterative way without complicated matrix inversion, which can reduce the overall complexity by one order of magnitude. The performance guarantee of the proposed GS-based precoding is analyzed from the following three aspects. At first,  we prove that GS-based precoding satisfies the transmit power constraint. Then, we prove that GS-based precoding enjoys a faster convergence rate than the recently proposed Neumann-based precoding. At last,  the convergence rate achieved by GS-based precoding is quantified, which reveals that GS-based precoding converges faster with the increasing number of BS antennas. To further accelerate the convergence rate and reduce the complexity, we propose a zone-based initial solution to GS-based precoding, which is much closer to the final solution than the traditional initial solution. Simulation results demonstrate that the proposed scheme outperforms Neumann-based precoding, and achieves the exact capacity-approaching performance of the classical linear precoding schemes with only a small number of iterations both in Rayleigh fading channels and spatially correlated channels.
\end{abstract}

\begin{keywords}
Massive MIMO, linear precoding, Gauss-Seidel (GS) method, low complexity.
\end{keywords}

\section{Introduction}\label{S1}

\IEEEPARstart Multiple-input multiple-output (MIMO) technology has been successfully integrated in a series of well established wireless communication standards, such as the 4th generation (4G) cellular system standard LTE-A~\cite{Dongwoon12}, wireless LAN standard IEEE 802.11n~\cite{Skordoulis08}, etc. It is also considered as a promising key technology for future wireless systems~\cite{federico14}. Unlike the traditional small-scale MIMO (e.g., at most 8 antennas in LTE-A), massive MIMO, which equips a very large number of antennas (e.g., 256 antennas or even more) at the base station (BS) to simultaneously serve multiple users, is recently proposed~\cite{marzetta10}. It has been theoretically proved that massive MIMO can achieve orders of increase in spectrum and energy efficiency simultaneously~\cite{ngo11}.

However, realizing the very attractive merits of massive MIMO in practice faces several challenging problems, one of which is the low-complexity precoding in the downlink~\cite{lu2013overview}. In order to shift the complicated processing of multi-user interference cancelation from the users to the BS, two categories of precoding techniques, i.e., nonlinear and linear precoding, have been proposed. The optimal nonlinear precoding technique is the dirty paper coding (DPC), which has been proved to be able to achieve the ideal channel capacity by subtracting the potential interferences before transmission~\cite{costa1983writing}. However, it is very difficult to be realized in massive MIMO systems due to the high complexity of successive encoding and decoding. To achieve the close-optimal capacity with reduced complexity, some other nonlinear precoding techniques, such as vector perturbation (VP) precoding~\cite{razi2010sum,mazrouei2012vector} and lattice-aided precoding~\cite{lee2010lattice,yang2012codebook}, have been proposed, but their complexity is still unaffordable when the dimension of the MIMO system is large or the modulation order is high~\cite{goldberger11} (e.g., 256 antennas at the BS with 64 QAM modulation). To make a trade-off between the capacity and complexity, one can resort to linear precoding techniques, which can also achieve the capacity-approaching performance in massive MIMO systems, where the channel matrix are asymptotically orthogonal~\cite{rusek13}. The simplest linear precoding scheme is match filter (MF) precoding, which can only achieve the satisfying capacity when the number of antennas at the BS tends to infinite, while zero forcing (ZF) precoding can enjoy a much better performance than MF precoding for a more realistic MIMO system when the number of BS antennas is not very large~\cite{rusek13}. However, ZF precoding involves unfavorable complicated matrix inversion whose complexity is cubic with respect to the number of users. Very recently, ZF precoding based on Neumann series approximation algorithm (which is called as Neumann-based precoding in this paper) was proposed in~\cite{prabhu13} to reduce the computational complexity, which is realized by converting the matrix inversion into a series of matrix-vector multiplications. However, only a marginal reduction of complexity can be achieved.

In this paper, we propose a capacity-approaching  linear precoding with low complexity based on the Gauss-Seidel (GS) method~\cite{bjorck1996numerical} for massive MIMO. Specifically, the contributions  can be summarized as follows:

(1) We propose GS-based precoding to precode the original signal for transmission in an iterative way without complicated matrix inversion, and prove that the proposed scheme satisfies the total transmit power constraint. The complexity analysis shows that the overall complexity can be reduced by one order of magnitude, and simulation results demonstrate that GS-based precoding can achieve the capacity-approaching performance with only a small number of iterations.

(2) By exploiting some special channel properties of massive MIMO systems, we prove that GS-based precoding  enjoys a faster convergence rate than the recently proposed Neumann-based precoding. We also derive a tight upper bound of the Frobenius norm of the iteration matrix of GS-based precoding, which equivalently indicates the lower bound of the convergence rate achieved by the proposed scheme. In addition, we show that GS-based precoding will converge faster as the number of BS antennas increases.

(3) To further accelerate the convergence rate, we propose a zone-based initial solution to GS-based precoding, which can utilize \emph{a prior} information of the final solution and therefore is closer to the final solution than the traditional zero-vector initial solution. That means when the number of iterations is limited, the zone-based initial solution will lead to a faster convergence rate.

The rest of the paper is organized as follows. Section~\ref{S2} briefly introduces the system model of massive MIMO. Section~\ref{S3} specifies the proposed GS-based precoding, together with the performance analysis. The simulation results of achieved channel capacity and the bit error rate (BER) performance are shown in Section~\ref{S4}. Finally, conclusions are drawn in Section~\ref{S5}.

{\it Notation}: Lower-case and upper-case boldface letters denote vectors and matrices, respectively; ${( \cdot )^T}$, ${( \cdot )^H}$, ${( \cdot )^{ - 1}}$,  ${\det ( \cdot )}$, and ${{\rm{tr}}( \cdot )}$ denote the transpose, conjugate transpose, inversion, determinant, and trace of a matrix, respectively; ${{\left\|  \cdot  \right\|_F}}$  and ${{\left\|  \cdot  \right\|_2}}$ denote the Frobenius norm of a matrix and the ${{l_2}}$-norm of a vector, respectively; ${\left|  \cdot  \right|}$  and ${{( \cdot )^ * }}$ denote the absolute and conjugate operators, respectively; ${{\mathop{\rm Re}\nolimits} \{  \cdot \} }$  and ${{\mathop{\rm Im}\nolimits} \{  \cdot \} }$ denote the real part and imaginary part of a complex number, respectively; Finally, ${{\bf{I}}_N}$ is the $ N \times N $  identity matrix.

\section{System Model}\label{S2}
As illustrated in Fig. 1, we consider a massive MIMO system employing ${N}$  antennas at the BS to simultaneously serve  ${K}$ scheduled single-antenna users~\cite{marzetta10}. In such system, we usually have ${N \gg K}$, e.g., ${N = 256}$  and ${K = 16}$ are considered in~\cite{Hoydis,dai13}. The received signal vector ${{\bf{y}} = {[{y_1}, \cdot  \cdot  \cdot ,{y_K}]^T}}$ containing the received signals for ${K}$ users can be represented as
\begin{equation}\label{eq1}
{\bf{y}} = \sqrt {{\rho _f}} {\bf{Ht}} + {\bf{n}},
\end{equation}
where ${{\rho _f}}$ is the signal-to-noise ratio (SNR), ${{{\bf{H}}} \in {\mathbb{C}^{K \times N}}}$  denotes the flat Rayleigh fading channel matrix whose entries follow ${{\cal CN}(0,1)}$, ${{\bf{n}} = {[{n_1}, \cdot  \cdot  \cdot ,{n_K}]^T}}$  presents the additive white Gaussian noise (AWGN) vector with independent and identically distributed (i.i.d.) complex Gaussian random variables with zero mean and unit variance, ${{\bf{t}}}$ denotes the ${N \times 1}$  signal vector for actual transmission after precoding, which is obtained by
\begin{equation}\label{eq2}
{\bf{t}} = {\bf{Ps}},
\end{equation}
where ${{\bf{P}}}$ is the ${N \times K}$  precoding matrix, ${{\bf{s}} = {[{s_1}, \cdot  \cdot  \cdot ,{s_K}]^T}}$  presents the original signal for all ${K}$  users to be transmitted. Note that the total transmit power is usually limited in massive MIMO systems, which requires the precoding matrix ${{\bf{P}}}$ satisfy the total transmit power constraint~\cite{couillet2011random}
\begin{equation}\label{eq3}
{\rm{tr}}\left( {{{\bf{P}}^H}{\bf{P}}} \right) \le K.
\end{equation}

\begin{figure}[tp]
\begin{center}
\hspace*{+5mm}\includegraphics[width=0.8\linewidth]{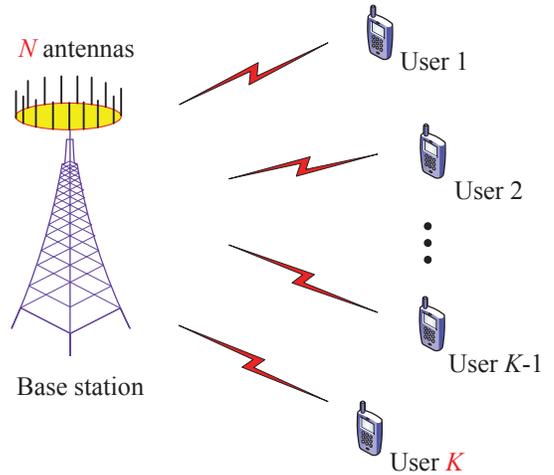}
\end{center}
\caption{System model of the massive MIMO system.} \label{FIG1}
\vspace{+3mm}
\end{figure}

The increased number of BS antennas ${N}$ (while the number of users ${K}$ is fixed) will lead to some special properties for massive MIMO. One attractive property is that the imperfect channel state information (CSI) is less probable to induce interferences to other users~\cite{kammoun2013linear}, which makes the linear precoding robust to  CSI mismatch. Another attractive property is that the columns of channel matrix ${{\bf{H}}}$ are asymptotically orthogonal as the scheduled users for communications are usually uncorrelated due to the physical separation~\cite{rusek13}. That means the simple linear precoding techniques can achieve the capacity-approaching performance with low complexity~\cite{rusek13}.

\section{Near-Optimal Linear Precoding With Low Complexity For Massive MIMO}\label{S3}
In this section, a low-complexity GS-based precoding is proposed to achieve the near-optimal performance without matrix inversion. We prove that GS-based precoding satisfies the total transmit power constraint. We also prove that it enjoys a faster convergence rate than the recently proposed Neumann-based precoding. To further accelerate the convergence rate, we propose a zone-based initial solution to GS-based precoding. Finally, the complexity analysis of the proposed scheme is provided to show its advantages over conventional schemes.

\subsection{Classical ZF precoding}\label{S2.1}
The classical linear ZF precoding is widely considered in massive MIMO systems due to it can eliminate multiuser interferences. The ZF precoding matrix can be presented as~\cite{rusek13}
\begin{equation}\label{eq4}
{{\bf{P}}_{{\rm{ZF}}}} = {\beta _{{\rm{ZF}}}}{{\bf{H}}^\dag } = {\beta _{{\rm{ZF}}}}{{\bf{H}}^H}{({\bf{H}}{{\bf{H}}^H})^{ - 1}} = {\beta _{{\rm{ZF}}}}{{\bf{H}}^H}{{\bf{W}}^{ - 1}},
\end{equation}
where ${{{\bf{H}}^\dag } = {{\bf{H}}^H}{({\bf{H}}{{\bf{H}}^H})^{ - 1}} = {{\bf{H}}^H}{{\bf{W}}^{ - 1}}}$  denotes the pseudo-inversion of the channel matrix ${{\bf{H}}}$ with  ${{\bf{W}}{\rm{ = }}{\bf{H}}{{\bf{H}}^H}}$,  ${\beta _{{\rm{ZF}}}}$ is the normalized factor which averages the fluctuations in transmit power. Substituting (\ref{eq4}) in (\ref{eq3}), we can obtain a suitable choice of ${\beta _{{\rm{ZF}}}}$ as
\begin{equation}\label{eq5}
{\beta _{{\rm{ZF}}}}  = \sqrt {\frac{K}{{{\rm{tr}}\left( {{{\bf{W}}^{ - 1}}} \right)}}}.
\end{equation}
Then, we can obtain the transmitted signal vector ${{\bf{t}}}$ after precoding as
\begin{equation}\label{eq6}
{\bf{t}} = {\beta _{{\rm{ZF}}}} {{\bf{H}}^H}{{\bf{W}}^{ - 1}}{\bf{s}} = {\beta _{{\rm{ZF}}}} {{\bf{H}}^H}{\bf{\hat s}},
\end{equation}
where we define ${{\bf{\hat s}} = {{\bf{W}}^{ - 1}}{\bf{s}}}$. Considering (\ref{eq1}) and (\ref{eq6}), we can use ${{\bf{G}} = {\bf{H}}}{{\bf{P}}_{{\rm{ZF}}}}$ to  present the equivalent channel matrix. Since the CSI is assumed to be known at the transmitter~\cite{prabhu13}, we have ${{\left| {{g_{mk}}} \right|^2} = 0}$  if  ${m \ne k}$, where ${{{g_{mk}}}}$ is the element of ${{\bf{G}}}$ in the ${m}$th row and ${k}$th column. So the received signal-to-interference-plus-noise ratio (SINR) ${\gamma _k}$ for any user ${k}$  can be computed as
\begin{equation}\label{eq7}
\begin{array}{l}
{\gamma _k} = \frac{{\frac{{{\rho _f}}}{K}{{\left| {{g_{kk}}} \right|}^2}}}{{\frac{{{\rho _f}}}{K}\sum\nolimits_{m \ne k}^K {{{\left| {{g_{mk}}} \right|}^2} + 1} }} = \frac{{{\rho _f}}}{K}{\left| {{g_{kk}}} \right|^2}\\
\\
\quad \; = \frac{{{\rho _f}}}{{{\rm{tr}}\left( {{{\bf{W}}^{ - 1}}} \right)}} \approx {\rho _f}\left( {\alpha  - 1} \right),
\end{array}
\end{equation}
where ${\alpha  = N/K}$. Note that the approximation in (7) is due to the fact that the matrix ${{{{\bf{W}}}}}$ is a Wishart matrix, and ${{{\rm{tr}}\left( {{{\bf{W}}^{ - 1}}} \right)}}$ is close to ${K/(N - K)}$ for a Wishart matrix of large dimension~\cite{tulino2004random}. Based on (\ref{eq7}), the sum rate achieved by ZF precoding can be presented as~\cite{rusek13}
\begin{equation}\label{eq8}
{C_{\rm ZF}} = \sum\nolimits_{i = 1}^K {{{\log }_2}(1 + {\gamma _i})}  \approx K{\log _2}\left( {1 + {\rho _f}(\alpha  - 1)} \right).
\end{equation}

Since ${\alpha  = N/K}$ is usually large in massive MIMO systems, from (\ref{eq8}) we can observe that ZF precoding can achieve the capacity close to the optimal DPC precoding~\cite{rusek13}. However, ZF precoding involves the matrix inversion ${{{\bf{W}}^{ - 1}}}$  of large size with the computational complexity ${{\cal O}({K^3})}$, which is high since ${K}$ is usually large in massive MIMO systems.

\subsection{Proposed GS-based Precoding}\label{S2.2}
Although the computation of ${{{\bf{W}}^{ - 1}}}$ is complicated, fortunately, the special channel property of massive MIMO enables us to obtain the precoded signal vector ${{\bf{t}}}$  (or equivalently ${{\bf{\hat s}}}$) in (\ref{eq6}) with low complexity.
For massive MIMO systems in the downlink, the columns of channel matrix ${{\bf{H}}}$  are asymptotically orthogonal~\cite{rusek13}. Therefore, we have  ${{{\bf{q}}^H}{\bf{Wq}} = {{\bf{q}}^H}{\bf{H}}{({{\bf{q}}^H}{\bf{H}})^H} > 0}$, where ${{\bf{q}}}$ is an arbitrary ${N \times 1}$ non-zero vector, which means the matrix ${{\bf{W}}}$ is positive definite. Besides, since we have  ${{{\bf{W}}^H} = {({\bf{H}}{{\bf{H}}^H})^H} = {\bf{W}}}$, we can conclude that the matrix ${{\bf{W}}}$  is Hermitian positive definite.

The special property that ${{\bf{W}}}$ is Hermitian positive definite in massive MIMO systems inspires us to exploit the GS method~\cite{bjorck1996numerical}, which can efficiently solve (\ref{eq6}) in an iterative way without matrix inversion, to obtain the precoded signal vector with low complexity. The GS method is used to solve the linear equation ${{\bf{Ax}} = {\bf{b}}}$, where ${{\bf{A}}}$ is the ${n \times n}$ Hermitian positive definite matrix, ${{\bf{x}}}$  is the ${n \times 1}$  solution vector, and ${{\bf{b}}}$  is the ${n \times 1}$  measurement vector. Unlike the traditional method that directly computes  ${{{\bf{A}}^{ - 1}}{\bf{b}}}$ to obtain ${{\bf{x}}}$, the GS method can iteratively solve the linear equation ${{\bf{Ax}} = {\bf{b}}}$ with low complexity. Since ${{\bf{W}}}$ is also Hermitian positive definite, we can decompose ${{\bf{W}}}$ as
\begin{equation}\label{eq9}
{\bf{W}} = {\bf{D}} + {\bf{L}} + {{\bf{L}}^H},
\end{equation}
where ${{\bf{D}}}$, ${{\bf{L}}}$, and ${{{\bf{L}}^H}}$  denote the diagonal component, the strictly lower triangular component, and the strictly upper triangular component of ${{\bf{W}}}$, respectively. Then we can exploit the GS method~\cite{bjorck1996numerical} to approximate ${{\bf{\hat s}} = {{\bf{W}}^{ - 1}}{\bf{s}}}$  in (\ref{eq6}) as below
\begin{equation}\label{eq10}
{{\bf{\hat s}}^{(i + 1)}} = {({\bf{D + L}})^{ - 1}}({\bf{s}} - {{\bf{L}}^H}{{\bf{\hat s}}^{(i)}}),\quad i = 0,1,2, \cdot  \cdot  \cdot
\end{equation}
where the superscript ${i}$ is the number of iterations, ${{{\bf{\hat s}}^{(0)}}}$ denotes the initial solution, which will be discussed later in Section~\ref{S3}-E, and the final solution ${{\bf{\hat s}}}$ satisfies ${{\bf{\hat s}} = \mathop {\lim }\limits_{i \to \infty } {{\bf{\hat s}}^{(i + 1)}}}$ and ${{\bf{\hat s}} = {({\bf{D}} + {\bf{L}})^{ - 1}}({\bf{s}} - {{\bf{L}}^H}{\bf{\hat s}})}$. According to (\ref{eq10}), the precoded signal vector for transmission can be achieved by
\begin{equation}\label{eq11}
{\bf{t}} = {\beta _{{\rm{GS}}}} {{\bf{H}}^H}{{\bf{\hat s}}^{(i + 1)}},
\end{equation}
where ${{\beta _{{\rm{GS}}}}}$ is the normalized factor of the proposed GS-based precoding, which will be discussed in detail in the next subsection. As ${({\bf{D + L}})}$ is a lower triangular matrix, one can solve (\ref{eq11}) to obtain ${{{\bf{\hat s}}^{(i + 1)}}}$  with low complexity as will be quantified in Section~\ref{S3}-F. It is worth noting that the proposed GS-based precoding is convergent for any initial solution since the matrix ${{\bf{W}}}$ is Hermitian positive definite~\cite[Theorem 7.2.2]{bjorck1996numerical}.

\subsection{Transmit power constraint of GS-based precoding}\label{S2.3}
The normalized factor ${{\beta _{{\rm{GS}}}}}$ in (11) is important to ensure GS-based precoding to satisfy the transmit power constraint. In this section, we will analyze how to select ${{\beta _{{\rm{GS}}}}}$ as below.

The GS method can be utilized to achieve the precoded signal vector ${{\bf{t}}}$ in an iterative way in (\ref{eq10}), and it can be also used to obtain the estimate of the matrix inversion ${{{\bf{W}}^{ - 1}}}$, which is necessary for the following analysis. Considering (\ref{eq10}), we can replace the ${K \times 1}$ vector ${{\bf{s}}}$ by an ${K \times K}$ identity matrix ${{{{\bf{I}}_K}}}$, and then the estimate ${{\bf{\hat W}}_{{\rm{inv}}}^{(i + 1)}}$ of ${{{\bf{W}}^{ - 1}}}$ in the (${i+1}$)th iteration can be acquired by
\begin{equation}\label{eq12}
{\bf{\hat W}}_{{\rm{inv}}}^{(i + 1)}\! =\! {\left( {{\bf{D}}\! +\! {\bf{L}}} \right)^{ - 1}}\left( {{{\bf{I}}_K}\! -\! {{\bf{L}}^H}{\bf{\hat W}}_{{\rm{inv}}}^{(i)}} \right),\quad i = 0,1, \cdot  \cdot  \cdot
\end{equation}
where ${{\bf{\hat W}}_{{\rm{inv}}}^{(0)}}$ is the initial solution. Since ${{{\bf{W}}^{ - 1}}}$ still satisfies ${{{\bf{W}}^{ - 1}} = {\left( {{\bf{D}} + {\bf{L}}} \right)^{ - 1}}\left( {{{\bf{I}}_K} - {{\bf{L}}^H}{{\bf{W}}^{ - 1}}} \right)}$, from (\ref{eq12}) we have
\begin{equation}\label{eq13}
{\bf{\hat W}}_{{\rm{inv}}}^{(i + 1)} - {{\bf{W}}^{ - 1}} = {\bf{B}}_{\rm GS}^{(i + 1)}\left( {{\bf{\hat W}}_{{\rm{inv}}}^{(0)} - {{\bf{W}}^{ - 1}}} \right),
\end{equation}
where ${{{\bf{B}}_{\rm GS}} =  - {\left( {{\bf{D}} + {\bf{L}}} \right)^{ - 1}}{{\bf{L}}^H}}$ denotes the iteration matrix of GS-based precoding. Based on (\ref{eq13}), the precoding matrix ${{{\bf{P}}_{\rm GS}}}$ can be presented as
\begin{equation}\label{eq14}
{{\bf{P}}_{\rm GS}} = {\beta _{\rm GS}}{{\bf{H}}^H}\left( {{\bf{B}}_{\rm GS}^{(i + 1)}\left( {{\bf{\hat W}}_{{\rm{inv}}}^{(0)} - {{\bf{W}}^{ - 1}}} \right) + {{\bf{W}}^{ - 1}}} \right).
\end{equation}
According to (\ref{eq3}), to meet the requirement of total transmit power constraint, ${{{\bf{P}}_{\rm GS}}}$ should satisfy the condition ${{\rm{tr}}\left( {{\bf{P}}_{\rm GS}^H{{\bf{P}}_{\rm GS}}} \right) \le K}$, which heavily depends on the choice of the normalized factor ${\beta _{\rm GS}}$. The following Lemma 1 and Lemma 2 will prove that we can choose ${{\beta _{\rm GS}} = {\beta _{\rm ZF}} = \sqrt {\frac{K}{{{\rm tr}({{\bf{W}}^{ - 1}})}}} }$ to satisfy the total transmit power constraint.

\vspace*{+2mm} \noindent\textbf{Lemma 1}. {\it For ${{\bf{{A}}} \in {\mathbb{C}^{n \times n}}}$, let ${{\lambda _{A,m}}}$ for ${m = 1, \cdot  \cdot  \cdot ,n}$ denotes the ${m}$th  eigenvalue of ${{\bf{A}}}$. If ${\left| {{\lambda _{A,m}}} \right| < 1}$, we have ${{\rm{tr}}\left( {{{\left( {{{\bf{A}}^H}{\bf{A}}} \right)}^k}} \right) < {\rm{tr}}\left( {{{\bf{A}}^k} + {{\left( {{{\bf{A}}^H}} \right)}^k}} \right)}$.}
\vspace*{+2mm}

\textit{Proof:} See Appendix A. \qed

\vspace*{+2mm} \noindent\textbf{Lemma 2}. {\it Let ${{{\bf{P}}_{\rm ZF}}}$ (${{\beta _{{\rm{ZF}}}}}$) and ${{{\bf{P}}_{\rm GS}}}$ (${{\beta _{{\rm{GS}}}}}$) denote the precoding matrix (normalized factor) of ZF precoding and GS-based precoding, respectively. In massive MIMO systems, if ${{\beta _{\rm GS}} = {\beta _{\rm ZF}}}$, we have ${{\rm{tr}}\left( {{\bf{P}}_{\rm GS}^H{{\bf{P}}_{\rm GS}}} \right){\rm{ < tr}}\left( {{\bf{P}}_{\rm ZF}^H{{\bf{P}}_{\rm ZF}}} \right)}$.}
\vspace*{+2mm}

\textit{Proof:} Based on (\ref{eq14}), the precoding matrix ${{{\bf{P}}_{\rm GS}}}$ can be presented as
\begin{equation}\label{eq15}
{{\bf{P}}_{\rm GS}} = {\beta _{\rm GS}}{{\bf{H}}^H}({{\bf{W}}^{ - 1}} + {\bf{E}}),
\end{equation}
where ${{\bf{E}} = {\bf{\hat W}}_{{\rm{inv}}}^{(i + 1)} - {{\bf{W}}^{ - 1}} = {\bf{B}}_{\rm GS}^{(i + 1)}({\bf{\hat W}}_{{\rm{inv}}}^{(0)} - {{\bf{W}}^{ - 1}})}$ is the error matrix between the estimate of ${{{\bf{W}}^{ - 1}}}$ and the real ${{{\bf{W}}^{ - 1}}}$. For simplicity but without loss of generality, we can choose ${{\bf{\hat W}}_{{\rm{inv}}}^{(0)}}$  as a zero matrix~\cite{bjorck1996numerical}, then the error matrix ${{\bf{E}}}$ will be ${{\bf{E}} = - {\bf{B}}_{\rm GS}^{(i + 1)}{{\bf{W}}^{ - 1}}}$. Substituting ${{\bf{E}} = - {\bf{B}}_{\rm GS}^{(i + 1)}{{\bf{W}}^{ - 1}}}$ into (\ref{eq15}), and using the fact that ${{\bf{W}} = {\bf{H}}{{\bf{H}}^H} = {{\bf{W}}^H}}$ and ${{\left( {{{\bf{W}}^{ - 1}}} \right)^H} = {{\bf{W}}^{ - 1}}}$, we have
\begin{equation}\label{eq16}
\begin{array}{l}
\frac{{{\bf{P}}_{\rm GS}^H{{\bf{P}}_{\rm GS}}}}{{\beta _{\rm GS}^2}} = ({{\bf{W}}^{ - 1}} + {{\bf{E}}^H}){\bf{H}}{{\bf{H}}^H}({{\bf{W}}^{ - 1}} + {\bf{E}})\\
\quad \quad \quad \; \, \, = {{\bf{W}}^{ - 1}} + {{\bf{E}}^H} + {\bf{E}} + {{\bf{E}}^H}{\bf{WE}}\\
\quad \quad \quad \; \, \, = {{\bf{W}}^{ - 1}}\! -\! {{\bf{W}}^{ - 1}}{\left( {{\bf{B}}_{\rm GS}^H} \right)^{(i + 1)}} \!-\! {\bf{B}}_{\rm GS}^{(i + 1)}{{\bf{W}}^{ - 1}}\\
\quad \quad \quad \quad \; \; + {{\bf{W}}^{ - 1}}{\left( {{\bf{B}}_{\rm GS}^H} \right)^{(i + 1)}}{\bf{WB}}_{\rm GS}^{(i + 1)}{{\bf{W}}^{ - 1}}.
\end{array}
\end{equation}
Since ${{{\bf{P}}_{\rm ZF}} = {\beta _{\rm ZF}}{{\bf{H}}^H}{{\bf{W}}^{ - 1}}}$ as shown in (\ref{eq4}), we can observe that the first term on the right side of (\ref{eq16}) is ${{{\bf{W}}^{ - 1}} = {\bf{P}}_{\rm ZF}^H{{\bf{P}}_{\rm ZF}}/\beta _{\rm ZF}^2}$. In addition, it is worth pointing out that for massive MIMO systems, ${{\bf{W}}}$ is a diagonal dominant matrix\cite{rusek13}. Therefore, matrix ${{{\bf{W}}^{ - 1}}}$ is also diagonal dominant. Then, we can utilize ${{{\bf{D}}^{ - 1}}}$ to reliably approximate ${{{\bf{W}}^{ - 1}}}$ and accordingly (\ref{eq16}) can be written as 
\begin{equation}\label{eq17}
\begin{array}{l}
\frac{{{\bf{P}}_{{\rm{GS}}}^H{{\bf{P}}_{{\rm{GS}}}}}}{{\beta _{{\rm{GS}}}^2}} = \frac{{{\bf{P}}_{{\rm{ZF}}}^H{{\bf{P}}_{{\rm{ZF}}}}}}{{\beta _{{\rm{ZF}}}^2}}\! -\! {{\bf{D}}^{ - 1}}{\left( {{\bf{B}}_{{\rm{GS}}}^H} \right)^{(i + 1)}}\! -\! {\bf{B}}_{{\rm{GS}}}^{(i + 1)}{{\bf{D}}^{ - 1}}\\
\quad \quad \quad \quad \; \; + {{\bf{D}}^{ - 1}}{\left( {{\bf{B}}_{{\rm{GS}}}^H} \right)^{(i + 1)}}{\bf{DB}}_{{\rm{GS}}}^{(i + 1)}{{\bf{D}}^{ - 1}}.
\end{array}
\end{equation}

In massive MIMO systems, the elements of the diagonal matrix ${{{\bf{D}}}}$ in (\ref{eq9}) can be well approximated by ${N}$ according to the random matrix theory~\cite{couillet2011random}, then the following expression can be obtained
\begin{equation}\label{eq18}
\begin{array}{l}
{\rm{tr}}\left( {\frac{{{\bf{P}}_{\rm GS}^H{{\bf{P}}_{\rm GS}}}}{{\beta _{\rm GS}^2}}} \right) = {\rm{tr}}\left( {\frac{{{\bf{P}}_{\rm ZF}^H{{\bf{P}}_{\rm ZF}}}}{{\beta _{\rm ZF}^2}}} \right) - \frac{1}{N} \cdot \\
\left[ {\rm{tr}}\left( {{{\left( {{\bf{B}}_{{\rm{GS}}}^H} \right)}^{(i + 1)}}\! +\! {\bf{B}}_{{\rm{GS}}}^{(i + 1)}} \right)\! -\! {\rm{tr}}\left( {{{\left( {{\bf{B}}_{{\rm{GS}}}^H{{\bf{B}}_{{\rm{GS}}}}} \right)}^{(i + 1)}}} \right) \right].
\end{array}
\end{equation}
Additionally, since ${{\bf{W}}}$ is a Hermitian positive definite matrix as we mentioned in Section III-A, we have~\cite[Theorem 7.2.2]{bjorck1996numerical}
\begin{equation}\label{eq19}
\mathop {\max }\limits_{1 \le m \le K} \left| {{\lambda _m}} \right| = \mathop {\max }\limits_{1 \le m \le K} \left| {\lambda _m^ * } \right| < 1,
\end{equation}
where ${{\lambda _m}}$ and ${\lambda _m^ * }$ are the eigenvalues of ${{{\bf{B}}_{\rm GS}}}$ and ${{\bf{B}}_{\rm GS}^H}$, respectively. Then, based on Lemma 1, if we replace ${{\bf{A}}}$ by ${{{\bf{B}}_{\rm GS}}}$, we have ${{\rm{tr}}\left( {{{\left( {{\bf{B}}_{{\rm{GS}}}^H{{\bf{B}}_{{\rm{GS}}}}} \right)}^{(i + 1)}}} \right) < {\rm{tr}}\left( {{{\left( {{\bf{B}}_{{\rm{GS}}}^H} \right)}^{(i + 1)}} + {\bf{B}}_{{\rm{GS}}}^{(i + 1)}} \right)}$, which means the second term on the right side of (\ref{eq18}) is large than zero. Thus, if we set ${{\beta _{\rm GS}} = {\beta _{\rm ZF}}}$, the following inequality is valid
\begin{equation}\label{eq20}
{\rm{tr}}\left( {{\bf{P}}_{\rm GS}^H{{\bf{P}}_{\rm GS}}} \right){\rm{ < tr}}\left( {{\bf{P}}_{\rm ZF}^H{{\bf{P}}_{\rm ZF}}} \right).
\end{equation}
\qed

\begin{figure}[tp]
\vspace{-3mm}
\begin{center}
\hspace*{-3mm}\includegraphics[width=1.1\linewidth]{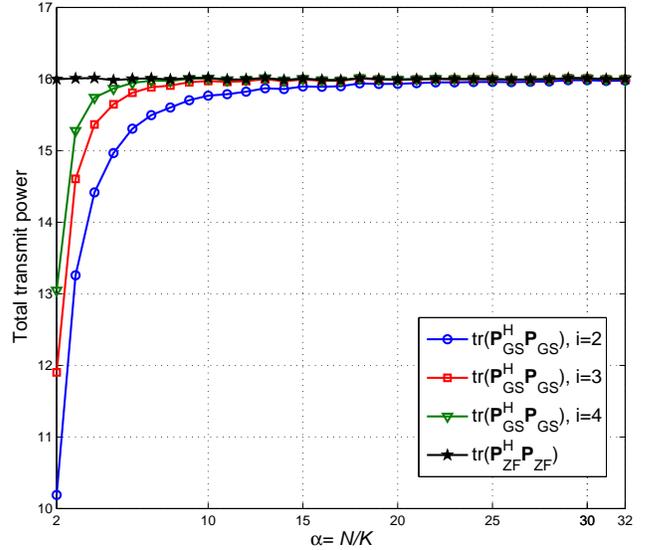}
\end{center}
\vspace{-4mm}
\caption{The total transmit power of GS-based precoding and ZF precoding when the number of users is fixed as ${K = 16}$ but the number of BS antennas ${N}$ is varying.} \label{FIG2}
\vspace{+4mm}
\end{figure}

Lemma 2 implies that when we choose ${{\beta _{\rm GS}} = {\beta _{\rm ZF}}}$, the total transmit power constraint ${{\rm{tr}}\left( {{\bf{P}}_{{\rm{GS}}}^H{{\bf{P}}_{{\rm{GS}}}}} \right) \le K}$ can be still satisfied since ${{\rm{tr}}\left( {{\bf{P}}_{{\rm{ZF}}}^H{{\bf{P}}_{{\rm{ZF}}}}} \right) \le K}$. What's more, as we can observe from (\ref{eq18}), the gap between ${{\rm{tr}}\left( {{\bf{P}}_{\rm GS}^H{{\bf{P}}_{\rm GS}}} \right)}$ and ${{\rm{tr}}\left( {{\bf{P}}_{\rm ZF}^H{{\bf{P}}_{\rm ZF}}} \right)}$ is negligible due to two facts: The first one is that for a large number of BS antennas ${N}$, the second term on the right side of (\ref{eq18}) is close to zero; The second one is that the proposed GS-based precoding is convergent, so the error matrix ${{\bf{E}}}$ (or equivalently ${{\bf{B}}_{{\rm{GS}}}^{(i + 1)}}$)  tends to be a zero matrix. Such small gap can be verified by Fig. 2, which shows the total transmit power of GS-based precoding and ZF precoding when ${K = 16}$ but ${N}$ is varying.

\subsection{Convergence rate}\label{S2.3}
In this part, we will prove that the proposed GS-based precoding can enjoy a faster convergence rate than the recently proposed Neumann-based precoding~\cite{prabhu13}.

From (\ref{eq10}), we can observe that the approximation error induced by GS-based precoding can be presented as
\begin{equation}\label{eq21}
{{\bf{\hat s}}^{(i + 1)}} - {\bf{\hat s}} = {{\bf{B}}_{\rm GS}}\left( {{{{\bf{\hat s}}}^{(i)}} - {\bf{\hat s}}} \right) =  \cdot  \cdot  \cdot  = {\bf{B}}_{\rm GS}^{(i + 1)}\left( {{{{\bf{\hat s}}}^{(0)}} - {\bf{\hat s}}} \right),
\end{equation}
where ${{{\bf{B}}_{\rm GS}} =  - {\left( {{\bf{D}} + {\bf{L}}} \right)^{ - 1}}{{\bf{L}}^H}}$ denotes the iteration matrix of GS-based precoding. Without loss of generality, we can utilize the ${{l_2}}$-norm to evaluate the approximation error as
\begin{equation}\label{eq22}
\begin{array}{l}
{\left\| {{{{\bf{\hat s}}}^{(i + 1)}} - {\bf{\hat s}}} \right\|_2} = {\left\| {{\bf{B}}_{\rm GS}^{(i + 1)}} \right\|_F}{\left\| {{{{\bf{\hat s}}}^{(0)}} - {\bf{\hat s}}} \right\|_2}\\
\quad \quad \quad \quad \quad \quad \le \left\| {{{\bf{B}}_{\rm GS}}} \right\|_F^{(i + 1)}{\left\| {{{{\bf{\hat s}}}^{(0)}} - {\bf{\hat s}}} \right\|_2},
\end{array}
\end{equation}
which indicates that the final approximation error of GS-based precoding is affected by two factors: the Frobenius norm of ${{{\bf{B}}_{\rm GS}}}$  and the ${{l_2}}$-norm (distance) between the initial solution ${{{\bf{\hat s}}^{(0)}}}$ and the final (exact) solution ${{\bf{\hat s}}}$.

Firstly, we will discuss the impact of ${{\left\| {{{\bf{B}}_{\rm GS}}} \right\|_F}}$ on the convergence rate. As we can observe from (\ref{eq22}), a smaller ${{\left\| {{{\bf{B}}_{\rm GS}}} \right\|_F}}$ will lead to a faster convergence rate~\cite{golub2012matrix}. The following Lemma 3 and Lemma 4 will verify that GS-based precoding enjoys a faster convergence rate than the recently proposed Neumann-based precoding~\cite{prabhu13}, since it has a smaller Frobenius norm of the iteration matrix.

\vspace*{+2mm} \noindent\textbf{Lemma 3}. {\it In massive MIMO systems, we have ${{\left\| {{{\bf{D}}^{ - 1}}{\bf{L}}} \right\|_p} < 1}$, where ${{\left\|  \cdot  \right\|_p}}$ is the ${{l_p}}$-norm of a matrix. }
\vspace*{+2mm}

\textit{Proof:} See Appendix B. \qed

\vspace*{+2mm} \noindent\textbf{Lemma 4}. {\it In massive MIMO systems, we have ${{\left\| {{{\bf{B}}_{\rm GS}}} \right\|_F} \le \frac{{{{\left\| {{{\bf{B}}_{\rm N}}} \right\|}_F}}}{{\sqrt 2 }}}$, where ${{{\bf{B}}_{\rm GS}} =  - {\left( {{\bf{D}} + {\bf{L}}} \right)^{ - 1}}{{\bf{L}}^H}}$  and ${{{\bf{B}}_{\rm N}} = {{\bf{D}}^{ - 1}}\left( {{\bf{L}} + {{\bf{L}}^H}} \right)}$  are the iteration matrices of GS-based precoding and Neumann-based precoding, respectively}.
\vspace*{+2mm}

\textit{Proof:} Note that  ${{{\bf{B}}_{\rm GS}}}$ can be rewritten as
\begin{equation}\label{eq23}
{{\bf{B}}_{\rm GS}}\! =\!  - {\left( {{\bf{D}}\! +\! {\bf{L}}} \right)^{ - 1}}{{\bf{L}}^H}\! =\!  - {({{\bf{I}}_K}\! +\! {{\bf{D}}^{ - 1}}{\bf{L}})^{ - 1}}{{\bf{D}}^{ - 1}}{{\bf{L}}^H}.
\end{equation}
Since ${{\left\| {{{\bf{D}}^{ - 1}}{\bf{L}}} \right\|_p} < 1}$  as we have proved in Lemma 3, the matrix ${{({{\bf{I}}_K} + {{\bf{D}}^{ - 1}}{\bf{L}})^{ - 1}}}$  can be expanded as~\cite[Theorem 2.2.3]{golub2012matrix}
\begin{equation}\label{eq24}
{({{\bf{I}}_K} + {{\bf{D}}^{ - 1}}{\bf{L}})^{ - 1}} = \sum\limits_{k = 0}^\infty  {{{( - 1)}^k}} {\left( {{{\bf{D}}^{ - 1}}{\bf{L}}} \right)^k}.
\end{equation}
In massive MIMO systems, where ${N \gg K}$, the elements of the diagonal matrix ${{{\bf{D}}}}$ in (\ref{eq9}) can be well approximated by ${N}$, while the elements of ${{\bf{L}}}$  will converge to 0, according to the random matrix theory~\cite{rusek13}. That means the term ${{\left( {{{\bf{D}}^{ - 1}}{\bf{L}}} \right)^k} \to 0}$ when ${k}$ is relatively large (e.g., ${k \ge 2}$). Therefore, we can approximate (\ref{eq24}) as ${{({{\bf{I}}_K} + {{\bf{D}}^{ - 1}}{\bf{L}})^{ - 1}} = {{\bf{I}}_K} - {{\bf{D}}^{ - 1}}{\bf{L}}}$ with a high accuracy, then the iteration matrix ${{{\bf{B}}_{\rm GS}}}$ of GS-based precoding can be approached by
\begin{equation}\label{eq25}
{{\bf{B}}_{\rm GS}} = - \left( {{{\bf{I}}_k} - {{\bf{D}}^{ - 1}}{\bf{L}}} \right){{\bf{D}}^{ - 1}}{{\bf{L}}^H} = {{\bf{D}}^{ - 1}}{\bf{L}}{{\bf{D}}^{ - 1}}{{\bf{L}}^H} - {{\bf{D}}^{ - 1}}{{\bf{L}}^H}.
\end{equation}
Then, the Frobenius norm of ${{{\bf{B}}_{\rm GS}}}$  can be calculated as
\begin{equation}\label{eq26}
\begin{array}{l}
{\left\| {{{\bf{B}}_{\rm GS}}} \right\|_F} = {\left\| {{{\bf{D}}^{ - 1}}{{\bf{L}}^H} - {{\bf{D}}^{ - 1}}{\bf{L}}{{\bf{D}}^{ - 1}}{{\bf{L}}^H}} \right\|_F}\\
\quad \quad \quad \; \; \; \le {\left\| {{{\bf{D}}^{ - 1}}{{\bf{L}}^H}} \right\|_F} + {\left\| {{{\bf{D}}^{ - 1}}{\bf{L}}{{\bf{D}}^{ - 1}}{{\bf{L}}^H}} \right\|_F}.
\end{array}
\end{equation}

Similar to the analysis above, the second term ${{\left\| {{{\bf{D}}^{ - 1}}{\bf{L}}{{\bf{D}}^{ - 1}}{{\bf{L}}^H}} \right\|_F}}$ on the right side of the inequality (\ref{eq26}) has a very limited contribution to ${{\left\| {{{\bf{B}}_{\rm GS}}} \right\|_F}}$, since the elements of the diagonal matrix ${{\bf{D}}}$ are large while the elements of ${{\bf{L}}}$ are close to zero. Thus, the second term ${{\left\| {{{\bf{D}}^{ - 1}}{\bf{L}}{{\bf{D}}^{ - 1}}{{\bf{L}}^H}} \right\|_F}}$ can be neglected and ${{\left\| {{{\bf{B}}_{\rm GS}}} \right\|_F}}$  can be upper bounded by
\begin{equation}\label{eq27}
{\left\| {{{\bf{B}}_{\rm GS}}} \right\|_F}\! \le \! {\left\| {{{\bf{D}}^{ - 1}}{{\bf{L}}^H}} \right\|_F}\! =\! {\left( {\sum\limits_{m = 1}^K {\sum\limits_{k = 1,k < m}^K {{{\left| {\frac{{{w_{mk}}}}{{{w_{mm}}}}} \right|}^2}} } } \right)^{\frac{1}{2}}},
\end{equation}
where ${{w_{mk}}}$ denotes the element of ${{\bf{W}} = {\bf{H}}{{\bf{H}}^H}}$ in the ${m}$th row and ${k}$th column.
Since the iteration matrix of Neumann-based precoding is ${{{\bf{B}}_{\rm N}} = {{\bf{D}}^{ - 1}}\left( {{\bf{L}} + {{\bf{L}}^H}} \right)}$~\cite{prabhu13}, the Frobenius norm of ${{{\bf{B}}_{\rm N}}}$  can be obtained by
\begin{equation}\label{eq28}
\begin{array}{l}
{\left\| {{{\bf{B}}_{\rm N}}} \right\|_F} = {\left( {\sum\limits_{m = 1}^K {\sum\limits_{k = 1,k \ne m}^K {{{\left| {\frac{{{w_{mk}}}}{{{w_{mm}}}}} \right|}^2}} } } \right)^{1/2}}\\
\quad \quad \quad \; \ = {\left( {2\sum\limits_{m = 1}^K {\sum\limits_{k = 1,k < m}^K {{{\left| {\frac{{{w_{mk}}}}{{{w_{mm}}}}} \right|}^2}} } } \right)^{1/2}}\\
\quad \quad \quad \; \ = \sqrt 2 {\left\| {{{\bf{D}}^{ - 1}}{{\bf{L}}^H}} \right\|_F}.
\end{array}
\end{equation}
Combing (\ref{eq27}) and (\ref{eq28}), we can conclude that ${{\left\| {{{\bf{B}}_{\rm GS}}} \right\|_F} \le \frac{{{{\left\| {{{\bf{B}}_{\rm N}}} \right\|}_F}}}{{\sqrt 2 }}}$.   \qed

Lemma 4 implies that GS-based precoding  enjoys an obviously faster convergence rate than Neumann-based precoding.  Furthermore, by exploiting some special channel properties of massive MIMO, we can derive a tight quantified upper bound of the Frobenius norm of  ${{{\bf{B}}_{\rm GS}}}$ to provide more insight of the convergence rate achieved by GS-based precoding as proved by the following Lemma 5.

\vspace*{+2mm} \noindent\textbf{Lemma 5}. {\it In massive MIMO systems, the Frobenius norm of the iteration matrix ${{{\bf{B}}_{\rm GS}}}$ of GS-based precoding is upper bounded by ${{\left\| {{{\bf{B}}_{\rm GS}}} \right\|_F} \le \sqrt {\frac{{{K^2} - K}}{{2N}}} }$}.
\vspace*{+2mm}

\textit{Proof:} Based on the conclusion of Lemma 4, we have ${\left\| {{{\bf{B}}_{\rm GS}}} \right\|_F^2 \le \frac{1}{2}\sum\limits_{m = 1}^K {\sum\limits_{k = 1,k \ne m}^K {{{\left| {\frac{{{w_{mk}}}}{{{w_{mm}}}}} \right|}^2}} } }$. Since in massive MIMO systems, the diagonal elements ${{w_{mm}}}$ of matrix ${{\bf{W}}}$  can be well approximated by ${N}$~\cite{rusek13}, we have the following inequity with probability one as
\begin{equation}\label{eq29}
\left\| {{{\bf{B}}_{\rm GS}}} \right\|_F^2 \le \frac{1}{{2{N^2}}}\sum\limits_{m = 1}^K {\sum\limits_{k = 1,k \ne m}^K {\left| {{(w_{mk})}^2} \right|} } .
\end{equation}

Considering the definition of the matrix ${{\bf{W}}}$ in (\ref{eq4}), we have ${{w_{mk}} = {\bf{h}}_m{{\bf{h}}_k^H} = \sum\limits_{l = 1}^N {{h_{ml}}h_{kl}^ * } }$  for ${m \ne k}$, where  ${{{\bf{h}}_m}}$ denote the  ${m}$th row of channel matrix ${{\bf{H}}}$, and ${{h_{ml}}}$ presents the element of ${{\bf{H}}}$  in the ${m}$th row and ${l}$th column. Then, we have
\begin{equation}\label{eq30}
\begin{array}{l}
{(w_{mk})}^2 = {\left( {\sum\limits_{l = 1}^N {{h_{ml}}{h_{kl}^*}} } \right)^2}\\
\quad \quad \;\; = \sum\limits_{l = 1}^N {{{\left( {{h_{ml}}h_{kl}^ * } \right)}^2}}  + 2\sum\limits_{l = 1,}^N {\sum\limits_{t = 1}^{l - 1} {{h_{ml}}h_{kl}^ * {h_{mt}}h_{kt}^ * } } .
\end{array}
\end{equation}

Note that all elements of channel matrix ${{\bf{H}}}$  are complex Gaussian random variables, then ${{h_{ml}}h_{kl}^ * }$  will follow the two-dimensional normal distribution whose jointly probability density function (p.d.f.) can be expressed as~\cite{williams1991probability}
\begin{equation}\label{eq31}
\begin{array}{l}
f\left( {{h_{ml}},h_{kl}^ * } \right) = \frac{1}{{2\pi {\sigma _{ml}}{\sigma _{kl}}\sqrt {1 - {\eta ^2}} }}\exp \left\{ { - \frac{1}{{2(1 - {\eta ^2})}} \cdot } \right.\\
\left. {\left[ {\frac{{{{\left( {{h_{ml}} - {\mu _{ml}}} \right)}^2}}}{{\sigma _{ml}^2}}\! -\! 2\eta \frac{{\left( {{h_{ml}} - {\mu _{ml}}} \right)\left( {h_{kl}^ *  - {\mu _{kl}}} \right)}}{{{\sigma _{ml}}{\sigma _{kl}}}}\! +\! \frac{{{{\left( {h_{kl}^ * \! -\! {\mu _{kl}}} \right)}^2}}}{{\sigma _{kl}^2}}} \right]} \right\},
\end{array}
\end{equation}
where ${\eta }$ is the correlation coefficient between ${{h_{ml}}}$ and ${h_{kl}^ * }$, ${{\mu _{ml}}}$ (${\sigma _{ml}^2}$)  and  ${{\mu _{kl}}}$ (${\sigma _{kl}^2}$) are the mean (variance) of ${{h_{ml}}}$  and  ${h_{kl}^ * }$, respectively. As the elements of channel matrix ${{\bf{H}}}$ are modeled as i.i.d. complex Gaussian random variables with zero mean and unit variance, which means  ${\eta  = 0}$, ${{\mu _{ml}} = {\mu _{mk}} = 0}$, and  ${\sigma _{ml}^2 = \sigma _{kl}^2 = 1}$, (\ref{eq31}) can be then simplified as
\begin{equation}\label{eq32}
f\left( {{h_{ml}},{h_{kl}^ *}} \right) = \frac{1}{{2\pi }}\exp \left\{ { - \frac{1}{2}\left( {h_{ml}^2 + h{{_{kl}^ * }^2}} \right)} \right\},
\end{equation}
which indicates that the mean and variance of the random variable ${{h_{ml}}h_{kl}^ * }$  for arbitrary  ${l = 1, \cdot  \cdot  \cdot ,N}$ is zero and one, respectively. Besides, since the two variables ${{h_{ml}}h_{kl}^ * }$  and ${{h_{mt}}h_{kt}^ * }$  for ${l \ne t}$  are also independent, the mean of ${{h_{ml}}h_{kl}^ * {h_{mt}}h_{kt}^ * }$  will be zero. Then, the sum of ${{\left( {{w_{mk}}} \right)^2}}$ in (\ref{eq30}) should be 
\begin{equation}\label{eq33}
\begin{array}{l}
\sum\limits_{m = 1}^K {\sum\limits_{k = 1,k \ne m}^K {\left| {{(w_{mk})}^2} \right|} } \\
 = \sum\limits_{m = 1}^K {\sum\limits_{k = 1,k \ne m}^K {\left| {\sum\limits_{l = 1}^N {{{\left( {{h_{lk}}h_{lm}^ * } \right)}^2}}  + 2\sum\limits_{l = 1,}^N {\sum\limits_{t = 1}^{l - 1} {{h_{lk}}h_{lm}^ * {h_{tk}}h_{tm}^ * } } } \right|} } \\
 = NK(K - 1),
\end{array}
\end{equation}
where we utilize the fact that ${\mathop {\lim }\limits_{L \to \infty } \frac{1}{L}\sum\limits_{l = 1}^L {{{\left( {{h_{ml}}h_{kl}^ * } \right)}^2}}  = 1}$  and  ${\mathop {\lim }\limits_{L \to \infty } \frac{1}{L}\sum\limits_{l = 1}^L {{h_{ml}}h_{kl}^ * {h_{mt}}h_{kt}^ * }  = 0}$ according to the law of large numbers~\cite{williams1991probability}, which is widely used to analyze the performance of massive MIMO systems. Substituting (\ref{eq33}) into (\ref{eq29}), we have
\begin{equation}\label{eq34}
\left\| {{{\bf{B}}_{\rm GS}}} \right\|_F^2 \le \frac{{{K^2} - K}}{{2N}},
\end{equation}
which verifies the conclusion of Lemma 5.   \qed

\begin{figure}[tp]
\vspace{-2mm}
\begin{center}
\hspace*{-4mm}\includegraphics[width=1.1\linewidth]{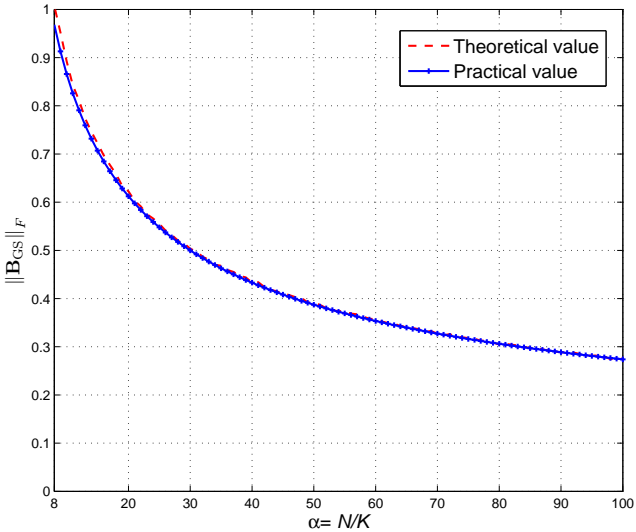}
\end{center}
\vspace{-3mm}
\caption{Comparison between the theoretical and practical ${{\left\| {{{\bf{B}}_{\rm GS}}} \right\|_F}}$  against ${\alpha  = N/K}$ (${K}$ is fixed as ${K = 16}$).} \label{FIG3}
\vspace{+4mm}
\end{figure}

Lemma 5 indicates that when the number of BS  antennas ${N}$ increases, while the number of users ${K}$  keeps fixed, the Frobenius norm of ${{{\bf{B}}_{\rm GS}}}$  will decrease, so a faster convergence rate can be achieved. This principle implies that the proposed GS-based precoding  is appropriate for massive MIMO systems.
Fig. 3 compares the theoretical and practical ${{\left\| {{{\bf{B}}_{\rm GS}}} \right\|_F}}$. We can observe that the gap between the theoretical and practical values is negligible, which means the derived upper bound of the Frobenius norms of ${{{\bf{B}}_{\rm GS}}}$  is tight for the proposed GS-based precoding in massive MIMO systems.

\subsection{Zone-based initial solution}\label{S2.3}
We now turn to discuss the selection of the initial solution ${{{\bf{\hat s}}^{(0)}}}$ in (\ref{eq10}), which plays an important role in the convergence rate as quantified in (\ref{eq22}). Traditionally, due to no priori information of the final solution is available, the initial solution ${{{\bf{\hat s}}^{(0)}}}$ in (\ref{eq10}) is set as a zero vector~\cite{bjorck1996numerical}, which is simple but usually far away  from the final solution. In this subsection, we propose a zone-based initial solution to achieve a faster convergence rate by exploiting another special property that the matrix ${{\bf{W}}}$ is diagonally dominant in massive MIMO systems~\cite{rusek13}.

Firstly, we convert the complex-valued channel model ${{\bf{H}}}$ in (1) into the corresponding real-valued channel model ${{{\bf{H}}_R}}$ as~\cite{rusek13}
\begin{equation}\label{eq35}
{{\bf{H}}_R} = {\left[ \begin{array}{l}
{\mathop{\rm Re}\nolimits} \{ {{\bf{H}}}\} \quad  - {\mathop{\rm Im}\nolimits} \{ {{\bf{H}}}\} \\
{\mathop{\rm Im}\nolimits} \{ {{\bf{H}}}\} \quad \,\,\,\,\,\;\,{\mathop{\rm Re}\nolimits} \{ {{\bf{H}}}\}
\end{array} \right]_{2K \times 2N}}.
\end{equation}
We then define ${{{\bf{W}}_R} = {{\bf{H}}_R}{\bf{H}}_R^H}$, ${{{\bf{s}}_R} = {[{\rm Re}({\bf{s}})\quad {\rm Im}({\bf{s}})]^T}}$,  ${{{\bf{\hat s}}_R} = {[{\rm Re}({\bf{\hat s}})\quad {\rm Im}({\bf{\hat s}})]^T}}$, and ${{\bf{\hat s}}_R^{(0)} = {[{\rm{Re}}({{\bf{\hat s}}^{(0)}})\;\;\;{\kern 1pt} {\rm{Im}}({{\bf{\hat s}}^{(0)}})]^T}}$ as the real-valued  ${{\bf{W}}}$, ${{\bf{s}}}$, ${{\bf{\hat s}}}$, and ${{{\bf{\hat s}}^{(0)}}}$, respectively. According to (\ref{eq6}), we have
\begin{equation}\label{eq36}
{{\bf{\hat s}}_R} = {\bf{W}}_R^{ - 1}{{\bf{s}}_R},
\end{equation}
then we can derive the following Corollary 1.

\vspace*{+2mm} \noindent\textbf{Corollary 1}. {\it In massive MIMO systems, we have ${{\hat s_{R,m}}{s_{R,m}} > 0}$, where ${{\hat s_{R,m}}}$  and ${{s_{R,m}}}$ denote the  ${m}$th element of ${{{\bf{\hat s}}_R}}$ and ${{{\bf{s}}_R}}$, respectively. }
\vspace*{+2mm}

\textit{Proof:} Let ${{w_{R,mk}^{ - 1}}}$ denote the element of ${{\bf{W}}_R^{ - 1}}$ in the ${m}$th row and ${k}$th column, then according to (\ref{eq36}) we have
\begin{equation}\label{eq37}
{\hat s_{R,m}}{s_{R,m}} = \left( {\sum\limits_{k = 1}^{2K} {w_{R,mk}^{ - 1}{s_{R,k}}} } \right){s_{R,m}}.
\end{equation}
Based on the fact that the matrix ${{\bf{W}}_R}$ (or equivalently ${{\bf{W}}}$) in massive MIMO systems is diagonally dominant~\cite{rusek13}, ${{\bf{W}}_R^{ - 1}}$  is also diagonally dominant since its diagonal elements are converges to ${1/N}$ and off-diagonal elements are close to zero. Moreover, all the values of ${s_{R,m}}$ for ${m = 1, \cdot  \cdot  \cdot ,2K}$ are small compared with ${N}$, since they are usually taken from a normalized real modulation constellation. Thus, (\ref{eq37}) can be approximated by
\begin{equation}\label{eq38}
\begin{array}{l}
{{\hat s}_{R,m}}{s_{R,m}} = \left( {\sum\limits_{k = 1}^{2K} {w_{R,mk}^{ - 1}{s_{R,k}}} } \right){s_{R,m}}\\
\quad \quad \quad \quad \; \; \approx w_{R,mm}^{ - 1}{s_{R,m}}{s_{R,m}} > 0.
\end{array}
\end{equation}  \qed

Corollary 1 implies that we can simply determine the sign of the precoded signal ${{\hat s_{R,m}}}$  according to the original signal ${{s_{R,m}}}$ to be transmitted, which is known at the BS. Based on this observation, we can further obtain \emph{a prior} information of the final solution ${{\hat s_{R,m}}}$, so a more suitable initial solution can be selected. Specifically, we can calculate
\begin{equation}\label{eq39}
{\bf{g}} = {{\bf{s}}_R} - {{\bf{W}}_R} \times {(\underbrace {z,z, \cdot  \cdot  \cdot ,z}_{2K})^T},
\end{equation}
where ${z}$ is a real constant, and then we can decide whether ${{\hat s_{R,m}}}$ is larger than ${z}$ or not according to ${{g_m}}$ (the ${m}$th element of ${{\bf{g}}}$). This inspires us come to the idea of the zone-based initial solution by selecting several different ${z}$'s to divide the potential range of the final solution ${{\hat s_{R,m}}}$ into multiple non-overlapped zones, then we can determine the initial solution belong to which specific zone according to \emph{a prior} information of the final solution, and consequently a faster convergence rate can be expected. Next we will discuss how to determine the potential range of the final solution ${{\hat s_{R,m}}}$.

Since the entries of the real channel matrix ${{{\bf{H}}_R}}$ are i.i.d., Gaussian random variables with zero mean and ${1/2}$ variance, according to the random matrix theory, when ${N \gg K}$, we have~\cite{rusek13}
\begin{equation}\label{eq40}
\frac{{{{\bf{W}}_R}}}{N} = \frac{{{{\bf{H}}_R}{\bf{H}}_R^H}}{N} \approx {{\bf{I}}_{2K}},
\end{equation}
which indicates that we can use ${N{{\bf{I}}_{2K}}}$ to approximate the matrix ${{{{\bf{W}}_R}}}$. Then (\ref{eq36}) can be also approximated as
\begin{equation}\label{eq41}
{{\bf{\hat s}}_R} \approx \frac{1}{N}{{\bf{s}}_R}.
\end{equation}
Note that the entries of ${{{\bf{s}}_R}}$ take values from a real constellation, e.g., when the normalized 64 QAM is considered, the potential value of the real-valued final solution ${{\hat s_{R,m}}}$ should be within the set ${\frac{1}{\gamma }\{- 7, - 5, - 3, - 1,}$  ${+ 1, + 3, + 5, + 7\} }$, where ${\gamma  = N\sqrt {42} }$ is used for power normalization.

After the potential range of the final solution ${{\hat s_{R,m}}}$ is determined, the proposed zone-based initial solution can be described in \textbf{Algorithm 1}.

\begin{algorithm}[h]
\caption{Zone-based initial solution} \KwIn{1) The number of non-overlapped zones
${Z}$;
\\\hspace*{+10.3mm} 2) The cardinality of the real constellation ${\left| Q \right|}$;
\\\hspace*{+10.3mm} 3) ${{s_{R,m}}}$, the ${m}$th element of ${{{\bf{s}}_R}}$}
\KwOut{${\hat s_m^{(0)}}$, the ${m}$th element of initial solution ${{{\bf{\hat s}}^{(0)}}}$.}
 \textbf{If} ${{s_{R,m}}} > 0$\
 \\\hspace*{+2.5mm} 1) Based on (\ref{eq41}), calculate $Z/2 - 1$ boundary values
 \\\hspace*{+7mm} $z = 2n\left| Q \right|/(\gamma Z),n = 1,2, \cdot  \cdot  \cdot ,Z/2 - 1$;
 \\\hspace*{+2.5mm} 2) Use (\ref{eq39}) to decide which zone ${{\hat s_{R,m}}}$  belongs to,
 \\\hspace*{+7mm} i.e., ${\hat s_{R,m}} \subset \left[ {2(n - 1)\left| Q \right|/(\gamma Z), 2n\left| Q \right|/(\gamma Z)} \right]$;
 \\\hspace*{+2.5mm} 3) Locate ${\hat s_{R,m}^{(0)}}$  at the center of this specific zone.

\textbf{else if} ${{s_{R,m}}} < 0$\
 \\\hspace*{+2.5mm} 1) Based on (\ref{eq41}), calculate $Z/2 - 1$ boundary values
 \\\hspace*{+7mm} $z = -2n\left| Q \right|/(\gamma Z),n = 1,2, \cdot  \cdot  \cdot ,Z/2 - 1$;
 \\\hspace*{+2.5mm} 2) Use (\ref{eq39}) to decide which zone ${{\hat s_{R,m}}}$  belongs to,
 \\\hspace*{+7mm} i.e., ${\hat s_{R,m}} \subset \left[ { - 2n\left| Q \right|/(\gamma Z), - 2(n - 1)\left| Q \right|/(\gamma Z)} \right]$;
 \\\hspace*{+2.5mm} 3) Locate ${\hat s_{R,m}^{(0)}}$  at the center of this specific zone.

\textbf{else}
 \\\hspace*{+2.5mm} ${\hat s_{R,m}^{(0)}=0}$.

\textbf{end}

\textbf{Solution}: ${\hat s_m^{(0)} = \hat s_{R,m}^{(0)} + i \cdot \hat s_{R,m + K}^{(0)}}$
\end{algorithm}

\begin{figure}[tp]
\vspace{+2mm}
\begin{center}
\hspace*{3mm}\includegraphics[width=0.9\linewidth]{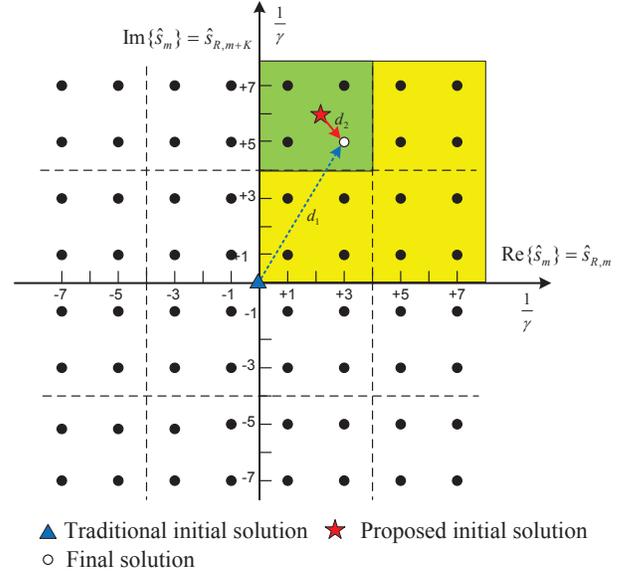}
\end{center}
\caption{Illustration of the zone-based initial solution to GS-based precoding, where 64 QAM and ${Z = 4}$ are considered as an example.} \label{FIG4}
\vspace{+4mm}
\end{figure}

Fig. 4 illustrated an example of \textbf{Algorithm 1} in a two-dimensional Cartesian coordinate system, where 64 QAM and ${Z = 4}$ are considered as an example. As ${{\hat s_{R,m}}}$ has 8 different values as we mentioned above, the cardinality of the real constellation is ${\left| Q \right| = 8}$. We first determine which quadrant ${{\bf{\hat s}}}$ belongs to by simply checking the sign of ${{s_{R,m}}}$ (the real part of ${{s_m}}$) and ${{s_{R,m+K}}}$ (the imaginary part of ${{s_m}}$), then we can divide the constellation into ${Z=4}$ zones to determine the initial solution, so the boundary values for different zones are ${z = \frac{1}{\gamma }\{  - 4, + 4\} }$. After that, we can calculate (\ref{eq39}) to judge which specific zone ${{\hat s_{m}}}$ belongs to, and finally locate the initial solution ${\hat s_m^{(0)}}$ at the center of this specific zone.

We can observe from Fig. 4 that the distance between the conventional zero-vector initial solution and the final solution is ${d{}_1}$, while the distance between the zone-based initial solution and the final solution is ${d{}_2}$, and obviously we have ${{d_2} < {d_1}}$. Thus, the proposed zone-based method provides a freedom of choice for the appropriate initial solution closer to the final solution ${{\bf{\hat s}}}$, which is crucial to ensure a faster convergence rate of GS-based precoding according to (\ref{eq22}), especially for high-order constellations like 64 QAM. Note that although the regular square QAM is considered in this paper, the idea of the proposed zone-based initial solution can be  extended to other modulations like APSK and other non-uniform constellations.

\subsection{Computational complexity analysis}\label{S2.4}
Since both ZF precoding and the proposed GS-based precoding need to compute ${{\bf{W}}\! =\! {\bf{H}}{{\bf{H}}^H}}$, we consider the computational complexity after the matrix ${\bf{W}}$ have  been obtained.  Besides, as the computational complexity is dominated by multiplications, we evaluate the complexity in terms of required number of complex multiplications. It can be found from (\ref{eq10}) and (\ref{eq11}) that the computational complexity of the proposed GS-based precoding comes from the following four parts.

1) The first one originates from solving the linear equation (\ref{eq10}). Considering the definition of ${{\bf{D}}}$ and ${{\bf{L}}}$ in (\ref{eq9}), the solution can be presented as
\begin{equation}\label{eq42}
\begin{array}{l}
\hat s_m^{(i + 1)}\! =\! \frac{1}{{{w_{mm}}}}({s_m} \!-\! \sum\limits_{k < m} {{w_{mk}}\hat s_k^{(i + 1)}\! -\! \sum\limits_{k > m} {{w_{mk}}\hat s_k^{(i)}} } ),\\
\quad \quad \quad \quad \quad \quad m,k = 1,2, \cdot  \cdot  \cdot ,K,
\end{array}
\end{equation}
where ${{\hat s_m}}$, ${\hat s_m^{(i + 1)}}$, and ${{s_m}}$  denote the  ${m}$th element of ${{{\bf{\hat s}}^{(i)}}}$, ${{{\bf{\hat s}}^{(i + 1)}}}$, and ${{\bf{s}}}$, respectively. It is clear that the required number of complex multiplications to compute ${\hat s_m^{(i{\rm{ + }}1)}}$  is  ${K}$. Since there are ${K}$  elements in  ${{{\bf{\hat s}}^{(i + 1)}}}$, solving the equation (\ref{eq10}) only requires ${{K^2}}$  times of complex multiplications.

2) The second one comes from the multiplication of a ${N \times K}$ matrix ${{{\bf{H}}^H}}$ and a ${K \times 1}$ vector ${{{\bf{\hat s}}^{(i + 1)}}}$, where ${NK}$ times of complex multiplications are required.

3) The third one is from the computation of the normalized factor ${{\beta _{{\rm{GS}}}}}$ in (\ref{eq11}), which can be simply chosen as ${{\beta _{{\rm{ZF}}}}}$ to satisfy the total transmit power constraint as we have proved in Section~\ref{S3}-C. It has been proved that when ${N}$ and ${K}$ go infinity while ${\alpha  = N/K}$  keeps fixed in massive MIMO systems, ${{\beta _{{\rm{ZF}}}}}$ converges to a deterministic value ${\sqrt {K(\alpha  - 1)} }$~\cite{rusek13}. Fig. 5 shows the comparison between the theoretical and practical ${{\beta _{{\rm{ZF}}}}}$ against different ${\alpha }$  when  ${K}$ is fixed to 16, where the practical ${{\beta _{{\rm{ZF}}}}}$ is obtained through intensive simulations. We can conclude from Fig. 5 that although ${N}$  and ${K}$  is finite in practical massive MIMO systems, the gap between the theoretical and practical ${{\beta _{{\rm{ZF}}}}}$  is negligible except when ${\alpha  = 1}$, while we usually have ${\alpha  > 1}$ in massive MIMO systems. Thus, once the system configuration has been fixed, the normalized factor ${{\beta _{{\rm{ZF}}}}}$ (or equivalently ${{\beta _{{\rm{GS}}}}}$) is known and constant, so we only need ${N}$  times of complex multiplications to compute ${{\beta _{{\rm{ZF}}}}{{\bf{H}}^H}{{\bf{\hat s}}^{(i + 1)}}}$.

4) The last one stems from the initial solution. From (\ref{eq39}), we can find that the zone-based initial solution only requires ${Z/2 - 1}$ (where ${Z}$ is the number of non-overlapped zones) times of multiplications of a ${2K \times 2K}$ real-valued matrix ${{\bf{W}}_R}$ and a  ${2K \times 1}$ real-valued vector. Note that all entries of the ${2K \times 1}$ vector are real constant ${z}$, so the multiplication can be realized by summing up each row of ${{\bf{W}}_R}$ and then multiplying the result by ${z}$. Thus, we only need ${(Z - 2)K}$ times of real multiplications to obtain the zone-based initial solution. As four real multiplications can be achieved by one complex multiplier, the complexity in this part is ${(Z - 2)K/4}$ times of complex multiplications. When ${Z = 4}$ is selected for 64 QAM as shown in Fig. 4, the required number of complex multiplications is as small as ${K/2}$.

\begin{figure}[tp]
\vspace{-3mm}
\begin{center}
\hspace*{-4mm}\includegraphics[width=1.1\linewidth]{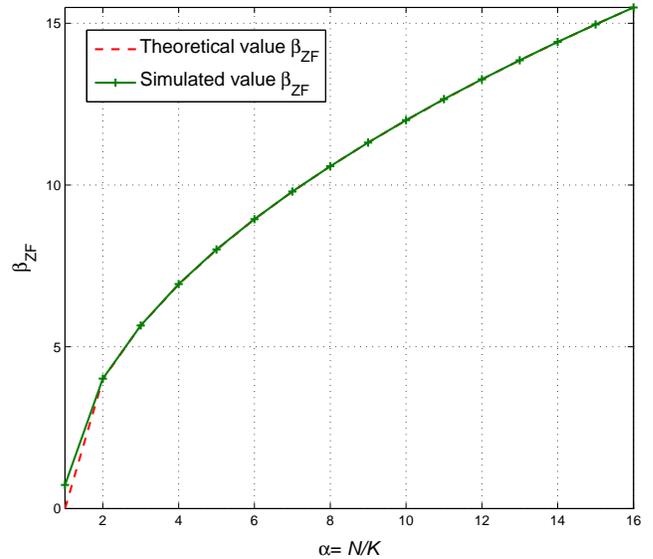}
\end{center}
\vspace{-2mm}
\caption{Comparison between the theoretical and practical ${{\beta _{{\rm{ZF}}}}}$ against ${\alpha  = N/K}$ (the number of users is fixed as ${K = 16}$).} \label{FIG5}
\end{figure}


\begin{figure}[tp]
\begin{center}
\hspace*{-4mm}\includegraphics[width=1.1\linewidth]{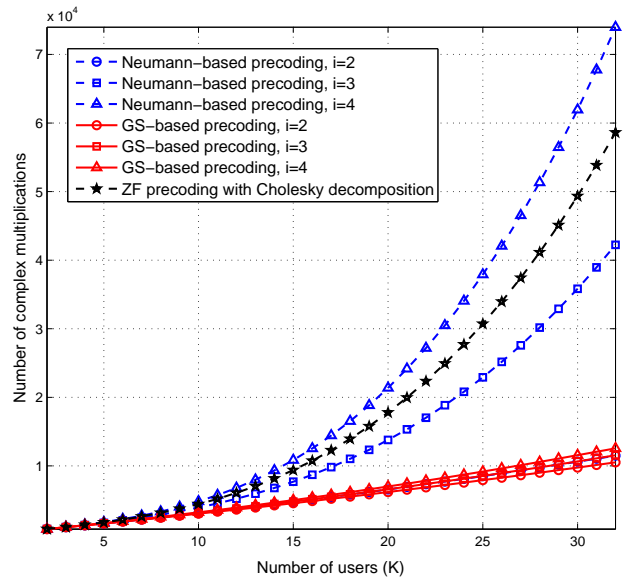}
\end{center}
\vspace{-4mm}
\caption{Complexity comparison against the number of users ${K}$ (the number of BS antennas is fixed as ${N=256}$).} \label{FIG6}
\vspace{+3mm}
\end{figure}

To sum up, the total number of multiplications required by the proposed GS-based precoding is ${N + NK + K/2 + i{K^2}}$. Fig. 6 compares the complexity of the recently proposed Neumann-based precoding~\cite{prabhu13} and the proposed GS-based precoding, whereby ZF precoding with Cholesky decomposition is also included as a baseline for comparison~\cite{Wu14}. We can observe from Fig. 6 that Neumann-based precoding has lower complexity than  ZF precoding with Cholesky decomposition when ${i \le 3}$, especially when ${i=2}$ with the complexity ${{\cal O}({K^2})}$. However, when ${i \ge 4}$, its complexity is still ${{\cal O}({K^3})}$, which is even higher than that of  ZF precoding. To ensure the approximation performance, usually a large value of  ${i}$ is required (as will be verified later in Section~\ref{S4}), which means the overall complexity of Neumann-based precoding is almost the same as  ZF precoding although it does not require any division operation which is difficult for hardware implementation~\cite{prabhu13}. In contrast, we can find that the proposed GS-based precoding does not involve any division operations either since ${\frac{1}{{{w_{mm}}}}}$ can be approximated by ${\frac{1}{N}}$~\cite{rusek13}, and its complexity is only ${{\cal O}({K^2})}$  for an arbitrary number of iterations. Even for ${i = 2}$, the proposed GS-based precoding has lower complexity  than Neumann-based precoding~\cite{prabhu13}.

Additionally, we can observe from (\ref{eq42}) that the computation of ${\hat s_m^{(i + 1)}}$ utilizes ${\hat s_k^{(i + 1)}}$ for ${k = 1,2, \cdot  \cdot  \cdot ,m - 1}$  in the current ${(i+1)}$th iteration and ${s_l^{(i)}}$ for ${l = m + 1,m + 2, \cdot  \cdot  \cdot ,K}$  in the previous ${i}$th iteration. Then two other benefits can be expected. Firstly, after ${\hat s_m^{(i + 1)}}$  has been obtained, we can use it to overwrite ${\hat s_m^{(i)}}$  which is useless in the next computation of ${\hat s_{m + 1}^{(i + 1)}}$. In this way, only one storage vector of size ${K \times 1}$  is required; Secondly, the solution to (10) becomes closer to the final solution ${{\bf{\hat s}}}$  with an increasing ${i}$, so ${\hat s_m^{(i + 1)}}$  can exploit the elements of ${\hat s_k^{(i + 1)}}$ for  ${k = 1,2, \cdot  \cdot  \cdot ,m - 1}$ that have already been computed in current ${(i+1)}$th iteration to produce more reliable result than Neumann-based precoding, which only utilizes all the elements of  ${{{\bf{\hat s}}^{(i)}}}$ in the previous ${i}$th iteration. Thus, the required number of iterations to achieve a certain approximation accuracy becomes smaller. Based on these two special advantages of the GS method, the overall complexity of the proposed GS-based precoding can be reduced further.

\section{Simulation Results}\label{S4}
To evaluate the performance of the proposed GS-based precoding, we provide the simulation results of the achievable channel capacity as well as the BER performance compared with the recently proposed Neumann-based precoding~\cite{prabhu13}. The capacity and BER performance of the classical  ZF precoding with Cholesky decomposition is also included as the benchmark for comparison. Besides, we also provide the performance of the optimal DPC precoding to verify the capacity-approaching performance of the proposed GS-based precoding. We consider two typical massive MIMO configurations with ${N \times K = 256 \times 16}$ and ${N \times K = 256 \times 32}$, respectively. The modulation scheme of 64 QAM is employed, and the SNR is defined at
the transmitter, i.e., ${{\rho _f}}$ in (\ref{eq1})~\cite{prabhu13}.

Firstly, we consider the uncorrelated Rayleigh fading channel as described in Section~\ref{S2}. Fig. 7 shows the capacity  comparison between Neumann-based precoding and GS-based precoding with zero-vector initial solution. The MIMO configuration is  ${N \times K = 256 \times 16}$, and  ${i}$ denotes the number of iterations. It is clear from Fig. 7 that the classical ZF precoding is capacity-approaching compared to the optimal DPC precoding, since their performance gap is within 0.5 dB for the achieved capacity of 220 bps/Hz. In addition, as shown in Fig. 7, when the number of iterations is small, e.g., ${i = 2}$, Neumann-based precoding cannot converge, leading to the serious multi-user interferences and the obvious loss in capacity, while the proposed GS-based precoding can achieve much better performance. For example, when SNR = 30 dB, the proposed scheme can achieve 175 bps/Hz, while only 130 bps/Hz can be achieved by Neumann-based precoding. As the number of iterations ${i}$ increases, the performance of both schemes improves. However, when the same number of iterations ${i}$  is used, the proposed scheme outperforms the Neumann-based one. For example, when  ${i = 4}$, the required SNR to achieve  the capacity of  200 bps/Hz by GS-based precoding is 26 dB, while Neumann-based precoding requires the SNR of 30 dB.

\begin{figure}[tp]
\vspace{-3mm}
\begin{center}
\hspace*{-4mm}\includegraphics[width=1.1\linewidth]{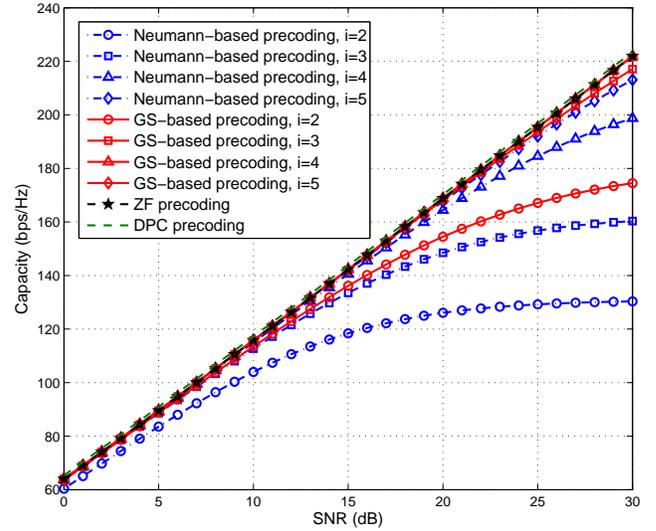}
\end{center}
\vspace{-4mm}
\caption{Capacity comparison between Neumann-based precoding and GS-based precoding with the zero-vector initial
solution for the ${256 \times 16}$  massive MIMO system.} \label{FIG7}
\vspace{-2mm}
\end{figure}

\begin{figure}[tp]
\begin{center}
\hspace*{-4mm}\includegraphics[width=1.1\linewidth]{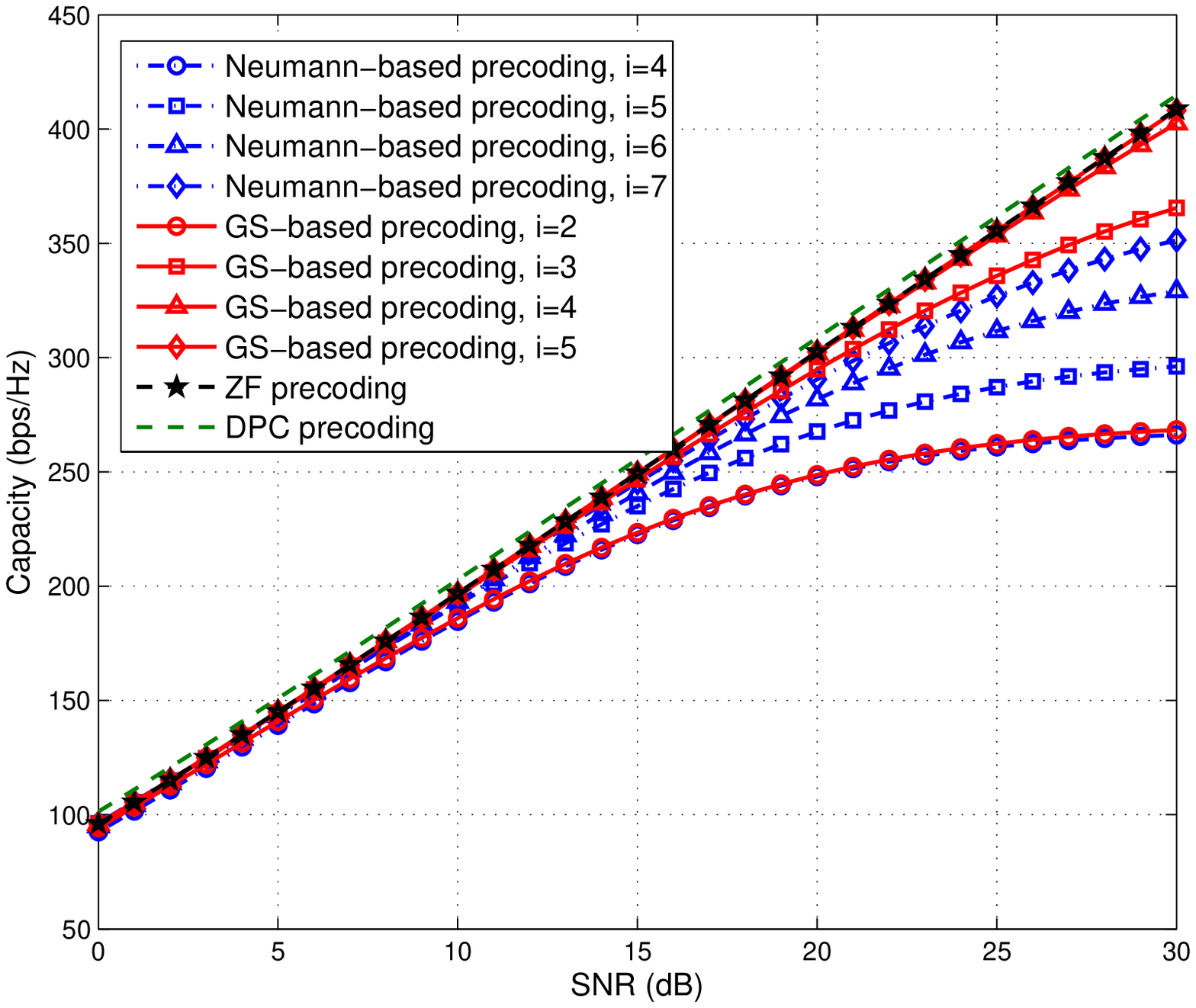}
\end{center}
\vspace{-4mm}
\caption{Capacity comparison between Neumann-based precoding and GS-based precoding with the zero-vector initial
solution for the ${256 \times 32}$  massive MIMO system.} \label{FIG8}
\vspace{+4mm}
\end{figure}

\begin{figure}[tp]
\setlength{\abovecaptionskip}{-10pt}
\setlength{\belowcaptionskip}{0pt}
\begin{center}
\hspace*{-4mm}\includegraphics[width=1.1\linewidth]{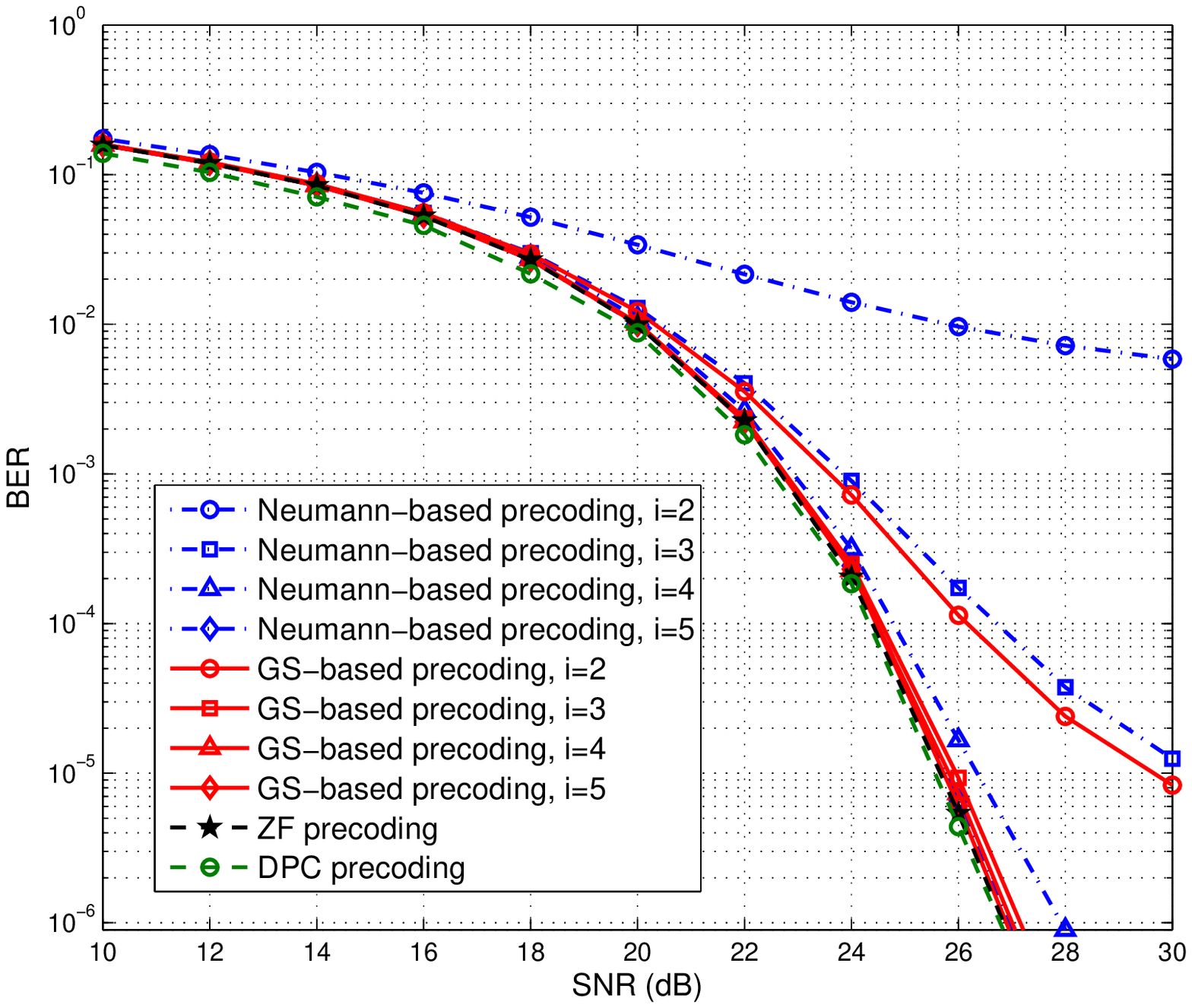}
\end{center}
\caption{BER performance comparison between Neumann-based precoding and GS-based precoding with the zero-vector initial
solution for the ${256 \times 16}$  massive MIMO system.} \label{FIG4}
\end{figure}

\begin{figure}[tp]
\setlength{\abovecaptionskip}{-10pt}
\setlength{\belowcaptionskip}{0pt}
\begin{center}
\hspace*{-4mm}\includegraphics[width=1.1\linewidth]{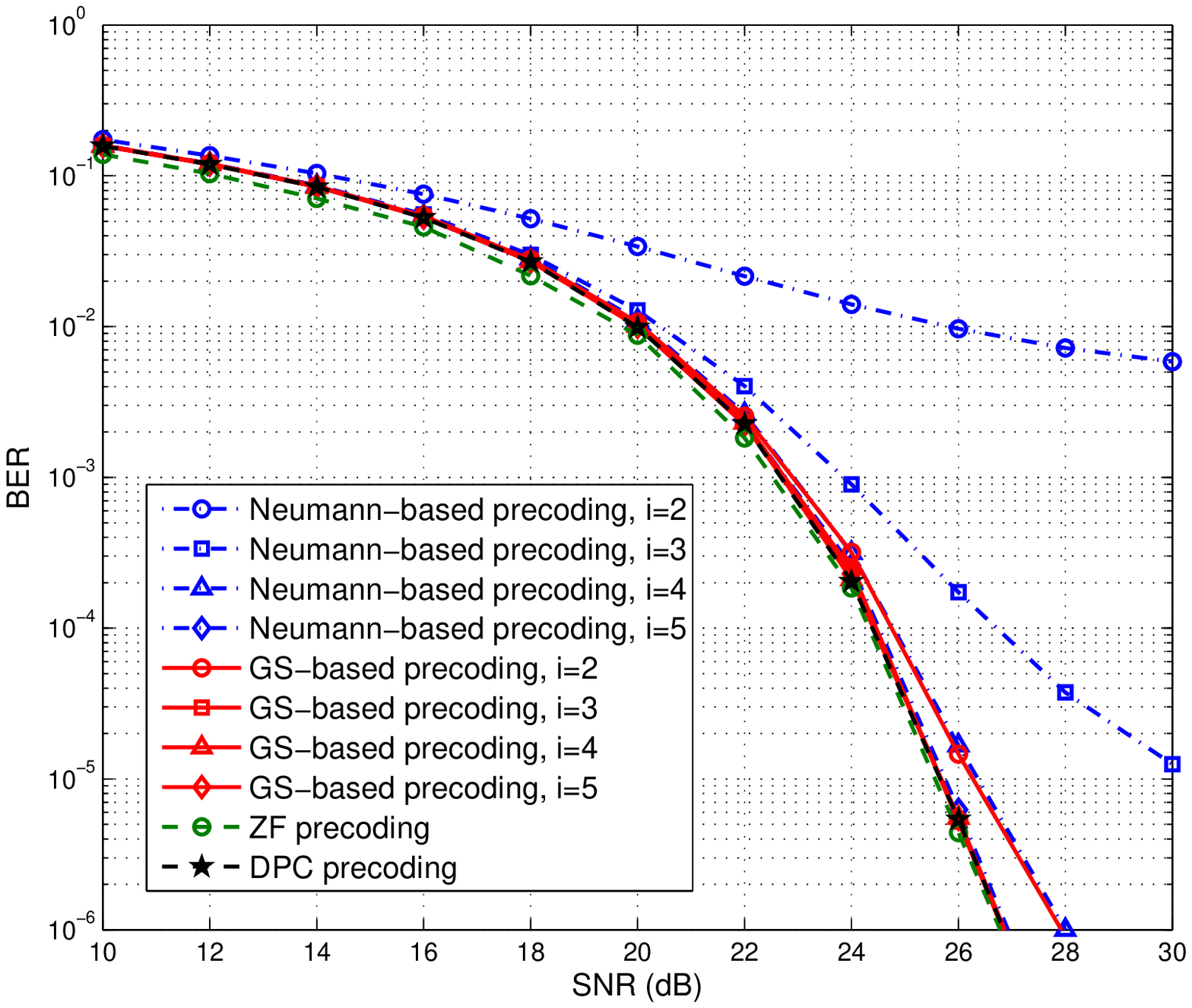}
\end{center}
\caption{BER performance comparison between Neumann-based precoding and GS-based precoding with the zone-based initial
solution for the ${256 \times 16}$ massive MIMO system.} \label{FIG6}
\end{figure}

\begin{figure}[tp]
\setlength{\abovecaptionskip}{-10pt}
\setlength{\belowcaptionskip}{0pt}
\begin{center}
\hspace*{-4mm}\includegraphics[width=1.1\linewidth]{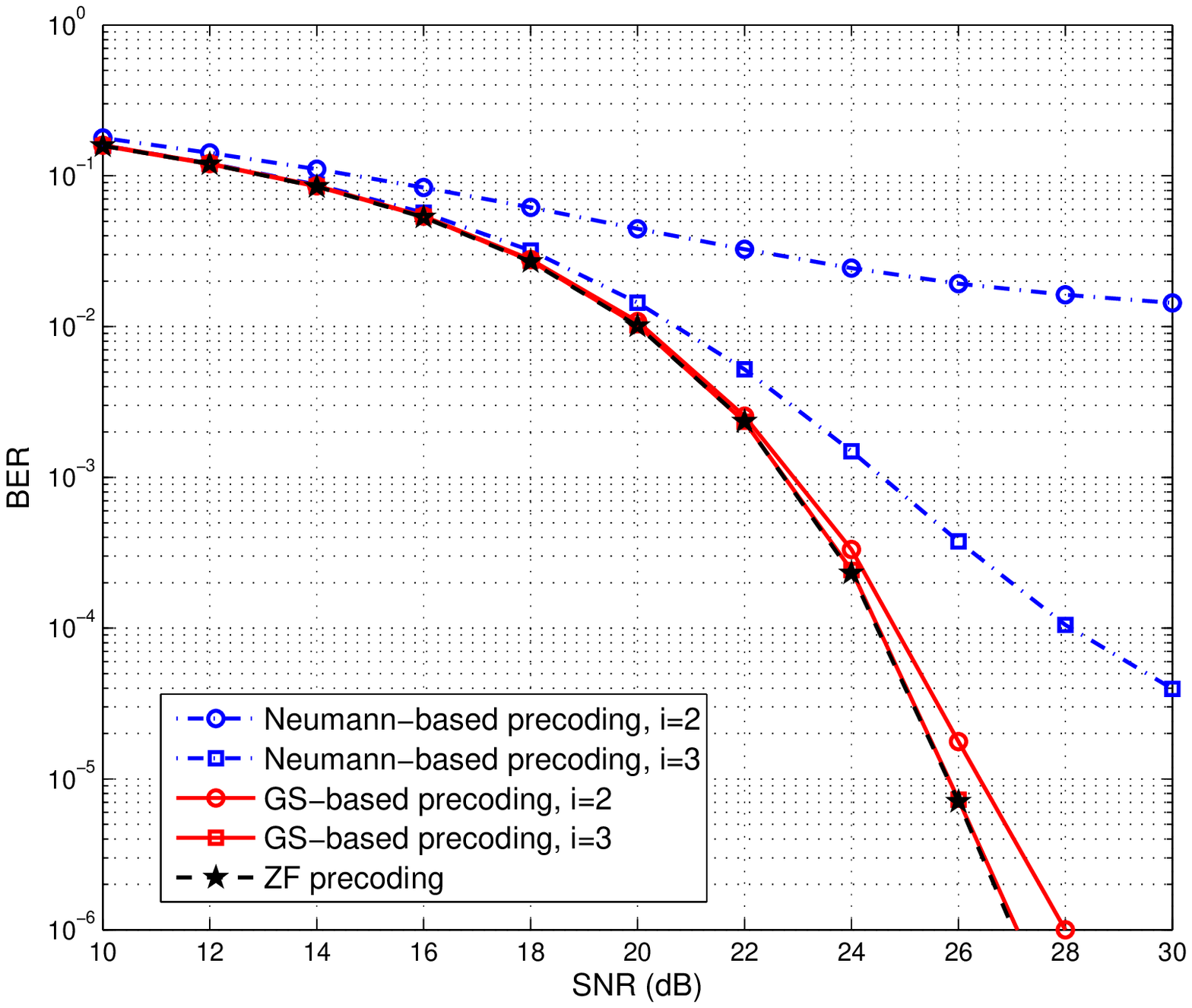}
\end{center}
\caption{BER performance comparison between Neumann-based precoding and GS-based precoding with the zone-based initial
solution for the ${256 \times 16}$ massive MIMO system with the antenna correlated factor ${\xi=0.2}$.} \label{FIG6}
\end{figure}

\begin{figure}[tp]
\setlength{\abovecaptionskip}{-10pt}
\setlength{\belowcaptionskip}{0pt}
\begin{center}
\hspace*{-4mm}\includegraphics[width=1.1\linewidth]{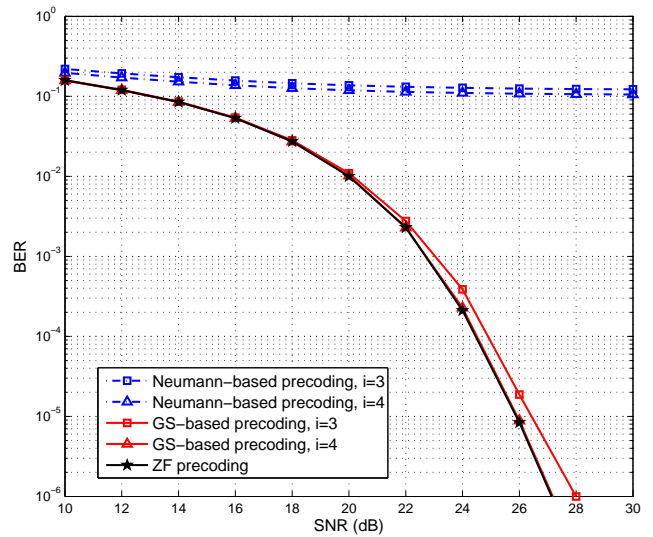}
\end{center}
\caption{BER performance comparison between Neumann-based precoding and GS-based precoding with the zone-based initial
solution for the ${256 \times 16}$ massive MIMO system with the antenna correlated factor ${\xi=0.5}$.} \label{FIG6}
\end{figure}

Fig. 8 shows the capacity comparison between Neumann-based precoding and GS-based precoding with zero-vector initial solution when ${N \times K = 256 \times 32}$. Comparing Fig. 7 and Fig. 8, we can find that with a decreasing value of ${\alpha  = N/K}$, the performance of Neumann-based precoding becomes worse. For example, when  ${i = 4}$, for the ${256 \times 16}$  MIMO system, Neumann-based precoding can achieve ${90\% }$ of capacity of the DPC precoding at SNR = 30 dB, while for the ${256 \times 32}$  MIMO system, it can only achieve ${64\% }$. In contrast, when ${i = 4}$, the proposed GS-based precoding can achieve ${99\% }$  and  ${97\% }$ of capacity of the DPC precoding for ${256 \times 16}$ and ${256 \times 32}$  MIMO systems, respectively. This indicates that the convergence of GS-based precoding is more robust with respect to the MIMO scales. Besides, given the same number of iterations, its superiority grows as the value of ${\alpha  = N/K}$ increases.

Fig. 9 and Fig. 10 show the BER performance comparison between Neumann-based precoding and GS-based precoding with different initial solutions, where ${N \times K = 256 \times 16}$ is considered. We can observe from Fig. 9 that the proposed GS-based precoding with the zero-vector initial solution requires a smaller number of iterations to obtain the same BER performance as Neumann-based precoding. When ${i = 2}$, the BER performance of the proposed GS-based precoding with zero-vector initial solution is almost the same as that of Neumann-based precoding when ${i = 3}$, which means a faster convergence rate can be achieved by the proposed scheme as we have proved in Section~\ref{S3}-D. Moreover, when the zone-based initial solution is used for GS-based precoding, we can observe from Fig. 10 that the convergence rate can be further accelerated. For example, to achieve the exact BER performance of ZF precoding, the required number of iterations by the proposed scheme with zero-vector initial solution is ${i = 4}$, while when the zone-based initial solution is used, the required number is reduced to ${i = 3}$. Therefore, the complexity can be further reduced. More importantly, we can find that when the number of iterations is relatively large (e.g., ${i = 3}$ in Fig. 10), the proposed GS-based precoding can achieve the near-optimal performance.

Finally, as the spatial correlation of MIMO channels plays a crucial role in the performance of realistic MIMO systems, we show in Fig. 11 and Fig. 12 how the channel correlation affects the performance of the proposed GS-based precoding. We adopt the exponential correlation channel model described in~\cite{Godana13}, where ${\xi }$ (${0 \le \xi  \le 1}$) denotes the antenna correlated factor between two adjacent antennas.
Comparing Fig. 11 and Fig. 12, we can observe that the performance of the classical ZF precoding degrades when the channel correlation becomes serious, which is consistent with the theoretical analysis in~\cite{Godana13}. It can be also observed that GS-based precoding can still converge to ZF precoding without obvious performance loss, while Neumann-based precoding can hardly converge due to the spatial correlation. However, the required number of iterations by the  proposed GS-based precoding to converge will becomes a little larger with an increasing value of ${\xi }$ (e.g., ${i = 3}$ when ${\xi=0.2}$ in Fig. 11, but ${i=4}$ when ${\xi=0.5}$ in Fig. 12), which means more serious channel correlation will lead to a slower convergence rate. Here, we only provide the results obtained by simulation, and the theoretical analysis of the impact of channel spatial correlation on the performance will be left for further study.  It is worth pointing out that the spatial correlation only lead to a slightly increased complexity for GS-based precoding, and the overall complexity is still much lower than Neumann-based precoding and ZF precoding with Cholesky decomposition.

\section{Conclusions}\label{S5}
In this paper, by fully exploiting some special channel property of massive MIMO systems, we have proposed a near-optimal linear precoding scheme with low complexity based on Gauss-Seidel method. The performance guarantee of the proposed GS-based precoding has been analyzed from the following three aspects. At first,  we proved that GS-based precoding satisfies the transmit power constraint. Then, we proved that GS-based precoding enjoys a faster convergence rate than the recently proposed Neumann-based precoding. At last,  the convergence rate achieved by GS-based precoding is quantified, which reveals that GS-based precoding converges faster with the increasing number of BS antennas. We have also proposed a zone-based initial solution to GS-based precoding to further accelerate the convergence rate. It is shown that the proposed scheme can reduce the complexity from ${{\cal O}({K^3})}$ to ${{\cal O}({K^2})}$. Simulation results demonstrate that the proposed scheme outperforms Neumann-based precoding, and approaches the optimal performance of the DPC precoding with a small number of iterations, e.g., ${i = 3}$ and ${i = 4}$ for an ${N \times K = 256 \times 16}$  massive MIMO system in the Rayleigh fading channel and spatially correlated channel with the antenna correlated factor ${\xi=0.5}$,  respectively.


\section*{Appendix A\\ Proof of Lemma 1}
According to the properties of the trace of a matrix~\cite{golub2012matrix}, we have
\begin{equation}\label{eq43}
\begin{array}{l}
{\rm{tr}}\left[ {{{\left( {{{\bf{A}}^H}{\bf{A}}} \right)}^k}} \right] \le \frac{1}{2}{\rm{tr}}\left[ {{{\left( {{{\bf{A}}^H}} \right)}^{2k}} + {{\bf{A}}^{2k}}} \right]\\
\quad \quad \quad \quad \quad \quad {\rm{ = }}\frac{1}{2}\left( {\sum\limits_{m = 1}^n {\lambda _{A,m}^{2k}}  + \sum\limits_{m = 1}^n {{{\left( {\lambda _{A,m}^ * } \right)}^{2k}}} } \right).
\end{array}
\end{equation}

Since ${\left| {{\lambda _{A,m}}} \right| < 1}$, we can conclude that
\begin{equation}\label{eq44}
\begin{array}{l}
{\rm{tr}}\left[ {{{\left( {{{\bf{A}}^H}{\bf{A}}} \right)}^k}} \right] \le \frac{1}{2}\left( {\sum\limits_{m = 1}^n {\lambda _{A,m}^{2k}}  + \sum\limits_{m = 1}^n {{{\left( {\lambda _{A,m}^ * } \right)}^{2k}}} } \right)\\
\quad \quad \quad \quad \quad \quad < \left( {\sum\limits_{m = 1}^n {\lambda _{A,m}^k}  + \sum\limits_{m = 1}^n {{{\left( {\lambda _{A,m}^ * } \right)}^k}} } \right)\\
\quad \quad \quad \quad \quad \quad = {\rm{tr}}\left[ {{{\bf{A}}^k} + {{\left( {{{\bf{A}}^H}} \right)}^k}} \right].
\end{array}
\end{equation}
\qed

\section*{Appendix B\\ Proof of Lemma 3}
According to the definition of the ${{l_p}}$-norm of a matrix, we have
\begin{equation}\label{eq45}
{\left\| {{{\bf{D}}^{ - 1}}{\bf{L}}} \right\|_p} = \mathop {\sup }\limits_{{\bf{x}} \ne 0} \frac{{{{\left\| {{{\bf{D}}^{ - 1}}{\bf{Lx}}} \right\|}_p}}}{{{{\left\| {\bf{x}} \right\|}_p}}},
\end{equation}
where ${{\bf{x}}}$ is an arbitrary ${K \times 1}$  nonzero vector. Since the matrix ${{\bf{L}}}$  is defined as the strictly lower triangular component of the matrix ${{\bf{W}}}$ (9), the following inequality can be guaranteed
\begin{equation}\label{eq46}
{\left\| {{{\bf{D}}^{ - 1}}{\bf{L}}} \right\|_p} = \mathop {\sup }\limits_{{\bf{x}} \ne 0} \frac{{{{\left\| {{{\bf{D}}^{ - 1}}{\bf{Lx}}} \right\|}_p}}}{{{{\left\| {\bf{x}} \right\|}_p}}} < \mathop {\sup }\limits_{{\bf{x}} \ne 0} \frac{{{{\left\| {{{\bf{D}}^{ - 1}}({\bf{L}} + {{\bf{L}}^H}){\bf{x}}} \right\|}_p}}}{{{{\left\| {\bf{x}} \right\|}_p}}}.
\end{equation}

Note that ${{\left\| {\bf{x}} \right\|_p} = {\left( {{{\left| {{x_1}} \right|}^p} +  \cdot  \cdot  \cdot  + {{\left| {{x_K}} \right|}^p}} \right)^{1/p}}}$, where ${{x_m}}$  denotes the  ${m}$th element of ${{\bf{x}}}$, and the numerator ${{{{\left\| {{{\bf{D}}^{ - 1}}({\bf{L}} + {{\bf{L}}^H}){\bf{x}}} \right\|}_p}}}$ can be presented as
\begin{equation}\label{eq47}
{\left( {{{\left| {{l_1}} \right|}^p} +  \cdot  \cdot  \cdot  + {{\left| {{l_K}} \right|}^p}} \right)^{1/p}},
\end{equation}
where
\begin{equation}\label{eq48}
\left| {{l_k}} \right| = \frac{{\left| {\sum\limits_{m = 1,m \ne k}^K {{w_{mk}}{x_m}} } \right|}}{{\left| {{w_{kk}}} \right|}}.
\end{equation}

In massive MIMO systems, where ${N \gg K}$, the matrix  ${{\bf{W}}}$ is diagonal dominant~\cite{rusek13}, which means the diagonal elements ${{w_{mm}}}$ is much larger than the off-diagonal elements ${{w_{mk}}}$  for ${m \ne k}$. Thus, we have ${\left| {{z_k}} \right| < \left| {{x_k}} \right|}$  for ${k = 1,2, \cdot  \cdot  \cdot ,K}$, based on which we can conclude that
\begin{equation}\label{eq49}
{\left\| {{{\bf{D}}^{ - 1}}{\bf{L}}} \right\|_p} < \mathop {\sup }\limits_{{\bf{x}} \ne 0} \frac{{{{\left\| {{{\bf{D}}^{ - 1}}({\bf{L}} + {{\bf{L}}^H}){\bf{x}}} \right\|}_p}}}{{{{\left\| {\bf{x}} \right\|}_p}}} < 1.
\end{equation}
\qed

\balance

\vspace*{+2mm}
\bibliography{IEEEabrv,Gao1Ref}

\end{document}